\documentclass[aip,jcp,amsmath,amssymb,floatfix,citeautoscript,longbibliography, reprint]{revtex4-1}

\usepackage[pdftex]{graphics}
\usepackage{subcaption}
\usepackage{tikz}
\usepackage{xr}
\usetikzlibrary{calc}
\usetikzlibrary{decorations.pathmorphing, patterns,shapes}
\usepackage{siunitx}
\usepackage{ragged2e}
\usepackage[labelsep=period,format=plain, justification=justified,font=footnotesize]{caption}
\usepackage[english]{babel}
\usepackage{float}
\usepackage{docmute} 
\makeatletter
\newif\ifsupplement
\supplementfalse
\def\maketitle{
  \@author@finish
  \ifsupplement
    \let\saved@title@column\title@column
    \let\title@column\relax
  \fi
  \title@column\titleblock@produce
  \suppressfloats[t]
  \ifsupplement
    \let\title@column\saved@title@column
  \fi}
\makeatother
\makeatletter
\newcommand{\suppressTOC}{
    \let\old@addcontentsline\addcontentsline
    \renewcommand{\addcontentsline}[3]{}}
\newcommand{\restoreTOC}{
    \let\addcontentsline\old@addcontentsline}
\makeatother
\newcommand{\beginsupplement}{
    \clearpage 
    \setcounter{table}{0} 
    \renewcommand{\thetable}{\arabic{table}} 
    \setcounter{figure}{0} 
    \renewcommand{\thefigure}{\arabic{figure}} 
    \setcounter{equation}{0} 
    \renewcommand{\theequation}{\arabic{equation}} 
    \setcounter{page}{1} 
    \renewcommand{\thepage}{\arabic{page}} 
    \setcounter{section}{0} 
    \renewcommand{\thesection}{\Roman{section}} 
    \restoreTOC 
    \supplementtrue 
    \onecolumngrid 
    \setcitestyle{numbers,square} 
}

\DeclareCaptionJustification{myjust}{\justifying}
\tikzset{snake it/.style={decorate, decoration={snake, amplitude=.5mm, segment length=2mm}}}
\tikzset{zigzag it/.style={decorate, decoration={zigzag, amplitude=.5mm, segment length=2mm}}}

\newcommand{\dd}{\mathrm{d}}

\newcommand{\cm}{\mathrm{cm}^{-1}}
\newcommand{\jw}{J(\omega)}

\newcommand{\ttw}{(\omega_3,t_2,t_1)}
\newcommand{\sttw}{S\ttw}

\newcommand{\wtw}{(\omega_3,t_2,\omega_1)}
\newcommand{\swtw}{S\wtw}
\newcommand{\weg}{\bar{\omega}_{eg}}
\newcommand{\ttt}{(t_3,t_2,t_1)}
\newcommand{\sttt}{S\ttt}
\newcommand{\tttau}{(\tau_3,\tau_2,\tau_1)}
\newcommand{\etmie}[2]{I^{[t_2=#1]}_{max}(#2)}

\newif\ifhighlightreviewchanges
\highlightreviewchangesfalse
\newcommand{\newchange}[1]{\ifhighlightreviewchanges \textcolor{red}{#1}\else #1\fi}

\newif\ifusealttext
\usealttextfalse
\newcommand{\alttext}[1]{\ifusealttext
\textit{Alt-Text: #1}\fi}

\begin{document}
\suppressTOC

\title{Streamlining Analysis and Design of Two-Dimensional Electronic Spectroscopy using Machine Learning}

\author{Nicholas I. Hausman}
\affiliation{Department of Chemistry, Stanford University, Stanford, California, 94305, USA}

\author{Joseph Kelly}
\affiliation{Department of Chemistry, Stanford University, Stanford, California, 94305, USA}

\author{Michael S. Chen}
\affiliation{Simons Center for Computational Physical Chemistry, Department of Chemistry, New York University, New York, New York 10003, USA}

\author{Frank Hu}
\affiliation{Department of Chemistry, Stanford University, Stanford, California, 94305, USA}

\author{Angela Lee}
\affiliation{Department of Chemistry, Massachusetts Institute of Technology, Cambridge, Massachusetts 02139, USA}

\author{Andr\'es Montoya-Castillo}
\affiliation{Department of Chemistry, University of Colorado Boulder, Boulder, Colorado, 80309, USA}

\author{Gabriela S. Schlau-Cohen}
\email{gssc@mit.edu}
\affiliation{Department of Chemistry, Massachusetts Institute of Technology, Cambridge, Massachusetts 02139, USA}

\author{Thomas E. Markland}
\email{tmarkland@stanford.edu}
\affiliation{Department of Chemistry, Stanford University, Stanford, California, 94305, USA}

\date{\today}

\begin{abstract}
Two-dimensional electronic spectroscopy (2DES) offers unique insights into the coupling between electronic and nuclear motion and dynamics, making it a key technique in diverse fields, including materials science and biology. Obtaining 2DES data requires a series of measurements that involve multiple pulses to construct the full picture — a time-consuming task that often necessitates working with limited or noisy data. Here we introduce a machine-learning based framework that aims to maximize the data that can be extracted from 2DES experiments and provides guidance towards the selection of additional experiments. We design a Gaussian mixture model to learn the underlying spectral density of a system, allowing the extraction of \newchange{reorganization energies} and the extrapolation of the 2DES spectra to other time delays beyond those measured, and demonstrate how our framework can be used to select additional measurements to further improve the accuracy. We show that our approach yields accurate results on a variety of systems, including simulations ranging from photoactive yellow protein in the gas phase to Nile red in benzene to the anionic green fluorescent protein chromophore in water, and experiments on Nile blue in ethanol. Our work provides an efficient route to extract maximum insights from 2DES while incurring minimal experimental costs.
\end{abstract}

\pacs{}

\maketitle
\section{Introduction}
\label{sec:intro}

Elucidating electronic dynamics and their modulation via nuclear motion is important to understand processes in systems such as light-harvesting complexes~\cite{mocaTwoDimensionalElectronicSpectroscopy2015}, nanomaterials~\cite{righettoDecipheringHotMultiexciton2018,collini2DElectronicSpectroscopic2021}, and solid-state photovoltaics~\cite{bolzonelloPhotocurrentDetected2DElectronic2021}. A key innovation for extracting this information is two-dimensional electronic spectroscopy (2DES)~\cite{hyblTwodimensionalElectronicSpectroscopy1998,jonasTwoDimensionalFemtosecondSpectroscopy2003}, an ultrafast nonlinear spectroscopic technique that provides a correlation map between excitation ($\omega_1$) and emission ($\omega_3$) frequencies separated by a delay time ($t_2$). The strengths of 2DES can be demonstrated when applied to condensed-phase systems, where the additional information revealed by 2D spectra can be used to disentangle features otherwise hidden in their linear counterparts. For example, in systems with many electronic states and/or many nuclear degrees of freedom, 2DES can be used to deduce couplings between electronic states, coherences, and relaxation pathways that are unresolved in linear absorption and emission spectroscopies and often obscured in transient absorption (TA). The utility of 2DES has motivated the development of methods for teasing out the additional information these spectra contain. In particular, relative populations of electronic states and their couplings can be obtained from frequency-frequency correlation maps, while other techniques also allow the extraction of quantities such as dynamic Stokes shifts~\cite{luExtractingFrequencyDependentDynamic2020}, coherences~\cite{mocaTwoDimensionalElectronicSpectroscopy2015,luExtractingFrequencyDependentDynamic2020,biswasCoherentTwoDimensionalBroadband2022,gajoTwoDimensionalElectronicSpectroscopy2025,milotaVibronicVibrationalCoherences2013}, and the timescales of relaxation dynamics~\cite{tokmakoffTwoDimensionalLineShapes2000,siemensResonanceLineshapesTwodimensional2010,bellAnalyticalCalculationTwodimensional2015}. However, extracting additional insights, particularly molecular parameters, from the available data presents an ongoing opportunity.

Machine learning (ML) has demonstrated considerable promise when applied to spectral analysis such as chiral discrimination of amino acids with terahertz spectroscopy~\cite{luoHighlyEfficientChiral2025}, identifying functional groups from FTIR~\cite{endersFunctionalGroupIdentification2021}, elucidating chemical structure from 1D NMR~\cite{huAccurateEfficientStructure2024}, facilitating analysis of 2D NMR spectra~\cite{liDEEPPickerDeep2021}, discerning secondary structures of proteins using 2DUV\cite{renMachineLearningRecognition2022} \newchange{or 2DIR}\cite{wuUnravelingDynamicProtein2024,yeAIProtocolRetrieving2025}, and also in predicting the spectra of molecules~\cite{renMachineLearningVibrational2021,mcnaughtonMachineLearningModels2023,carboneMachineLearningXRayAbsorption2020,paruzzoChemicalShiftsMolecular2018,gerrardIMPRESSIONPredictionNMR2020}. Two classes of ML applications to 2DES have emerged: the former of these has focused on using machine learned potential energy and electronic energy gap surfaces to enable efficient atomistic molecular dynamics simulations from which 2DES spectra are computed to gain insight into the underlying molecular processes~\cite{chenExploitingMachineLearning2020,kellyTwoDimensionalElectronicSpectroscopy2025,chenElucidatingRoleHydrogen2023}, while the latter has focused on developing methods that use ML to directly extract additional information from the spectra themselves.~\cite{rodriguezMachineLearningTwodimensional2019,namuduriMachineLearningEnabled2020,schultzUsingMachineLearning2025,parkerMappingSimulatedTwoDimensional2022,sbaitiMachineLearningVideo2025} Here we focus on the second of these by introducing a ML-based method to extract information and guide successive measurements.

Several ML architectures have previously been employed to extract information from 2DES data. Long short term memory (LSTM) frameworks, neural networks (NN), and convolutional neural networks (CNN) have been used to predict FMO dipole moment orientations from simulated spectra,~\cite{rodriguezMachineLearningTwodimensional2019} demonstrating how spectra-dependent molecular properties can be extracted from 2DES using ML. CNNs have been used to obtain homogeneous and inhomogeneous linewidths from simulated spectra and applied to experimental spectra of potassium atomic vapor and a quantum well~\cite{namuduriMachineLearningEnabled2020}, providing insights into relaxation dynamics. NN approaches have been developed to obtain Coulombic coupling strengths for simulated homodimers, suggesting that experimental conditions different from those typically used for human interpretation may improve ML performance~\cite{schultzUsingMachineLearning2025}. Although all of these methods use spectra at a single $t_2$ time delay, additional information is encoded in the spectral evolution across multiple $t_2$ delays. A recent ML approach used this information by adapting a video classification CNN to exploit multiple 2DES time delays to classify Coulombic coupling strengths of simulated homodimers into different regimes, while providing insight into key spectral regions for this task.~\cite{sbaitiMachineLearningVideo2025}

Here, we introduce a machine-learning based framework in the form of a Gaussian mixture model (GMM) that uses a minimal amount of 2DES information to extrapolate 2DES spectra forwards and backwards in time, predict other optical spectra, including linear absorption and pump-probe, and provide access to other important physical quantities such as the spectral density, vibronic couplings, and the reorganization energy. We demonstrate the effectiveness of our approach using recent 2DES experiments of Nile blue in ethanol and a wide range of simulated systems, including the photoactive yellow protein (PYP) in the gas phase~\cite{isbornElectronicAbsorptionSpectra2012}, Nile red in benzene~\cite{zuehlsdorffCombiningEnsembleFranckCondon2018}, the anionic green fluorescent protein (GFP) chromophore in water,~\cite{zuehlsdorffCombiningEnsembleFranckCondon2018} and Nile blue in ethanol~\cite{kellyTwoDimensionalElectronicSpectroscopy2025}. In each of these cases, we show that our ML-based framework accurately captures the spectral density of each system, as well as the corresponding linear absorption spectrum and 2DES spectra. We show how one can obtain high accuracy using only a single 2DES time delay ($t_2$), and how these predictions can be refined by including additional time delays. We illustrate how one can use an active learning strategy based on query by committee to guide the selection of additional time delays that provide the most information. By demonstrating our framework's performance on the experimental 2DES data of Nile blue in ethanol we illustrate its robustness to experimental conditions and hence its potential to extract important parameters in a data-efficient manner from a wide range of chemical and biological systems.

\section{Methods}
\label{sec:methods}

In this section, we first outline the theoretical background required to explain the approach we employ: using the second-order cumulant expansion of the energy-gap operator to approximate the 2DES in terms of the spectral density. We then discuss how we include experimental considerations, such as finite-width laser pulses and phasing. Finally, we show how a GMM framework can be used to construct a model for the spectral density that can be used to predict the 2DES and other optical properties.

\subsection{Electronic Spectroscopy Theory}
\label{sec:methods-theory}

2DES involves three light-matter interactions resulting in a signal proportional to the third-order polarization, $P^{(3)}$. Within linear response theory, this is given by~\cite{mukamelPrinciplesNonlinearOptical1995}
\begin{align}
    P^{(3)}(\mathbf{r},t) &= \int_0^\infty\dd \tau_3\int_0^\infty\dd \tau_2\int_0^\infty\dd \tau_1 \quad R\tttau\times \nonumber \\
    &\quad E(\mathbf{r},t-\tau_3)E(\mathbf{r},t-\tau_3-\tau_2) \times \nonumber \\ &\quad E(\mathbf{r},t-\tau_3-\tau_2-\tau_1),
    \label{eq:polarization}
\end{align}
where the electric field of the laser pulses $E(\mathbf{r},t)$ are convolved with the response function $R\tttau$. When the electric field arises from a sequence of pulses with time delays centered at $t+t_1+t_2$, $t+t_2$, and $t$, it can be expressed as a sum of the complex pulse fields~\cite{biswasCoherentTwoDimensionalBroadband2022,brixnerPhasestabilizedTwodimensionalElectronic2004}
\begin{align}
\label{eq:e_fields}
    E(\mathbf{r},t)&=E_1(\mathbf{r},t+t_1+t_2) + E_2(\mathbf{r},t+t_2) + E_3(\mathbf{r},t) \nonumber \\
&\quad+E_1^*(\mathbf{r},t+t_1+t_2) + E_2^*(\mathbf{r},t+t_2) + E_3^*(\mathbf{r},t)
\end{align}
for the two pump pulses $E_1$ and $E_2$ and the probe pulse $E_3$, where each can be expressed as~\cite{biswasCoherentTwoDimensionalBroadband2022,brixnerPhasestabilizedTwodimensionalElectronic2004}
\begin{align}
    E_n(\mathbf{r},t)=A_n(t)e^{-i\tilde{\omega}_nt+i\mathbf{k_n r}}.
\end{align}
Here $A_n(t)$ is the complex pulse envelope function, $\mathbf{k_n}$ is the wavevector, and $\tilde{\omega}_n$ is the carrier frequency. Upon enforcing time ordering of the pulses ($E_1$ followed by $E_2$ followed by $E_3$) and invoking the rotating wave approximation, substitution of Eq.~\ref{eq:e_fields} into Eq.~\ref{eq:polarization} yields the measured signal $\sttt$. The measured signal is a sum of eight terms involving the convolution of the response function with the three electric fields and their complex conjugates (SI Sec.~\ref{si:sig_eqs}).~\cite{mukamelPrinciplesNonlinearOptical1995,biswasCoherentTwoDimensionalBroadband2022,jonasTwoDimensionalFemtosecondSpectroscopy2003} For example, the rephasing (RP) signal is
\begin{align}
\label{eq:s_rp}
    S_\mathrm{RP}\ttt&\propto\int_0^\infty\dd\tau_3\int_0^\infty\dd\tau_2\int_0^\infty\dd\tau_1R\tttau\times \nonumber \\
    &\quad E_3(\mathbf{r},t_3-\tau_3)E_2(\mathbf{r},t_3+t_2-\tau_3-\tau_2)\times \nonumber \\
    &\quad E_1^*(\mathbf{r},t_3+t_1+t_2-\tau_3-\tau_2-\tau_1)
\end{align}
which is the signal detected when the phase matching condition $\mathbf{k = -k_1+k_2+k_3}$ is met~\cite{biswasCoherentTwoDimensionalBroadband2022,choTwoDimensionalOpticalSpectroscopy2009}. Experimentally, the signal can be detected using an interferometer that Fourier transforms the signal $\sttt$ from the $t_3$ time domain to the $\omega_3$ frequency domain.\cite{sonUltrabroadband2DElectronic2017} Additionally, a Fourier transform from $t_1$ to $\omega_1$ is often used to create frequency-frequency correlation maps $\swtw$ at various $t_2$ time delays for analysis.

The convolution of the response function $R\tttau$ with the pulse electric fields plays an important role in experimental 2DES where $E_n(\mathbf{r},t)$ has finite width and duration, which we discuss in Sec.~\ref{sec:exp2des}. In our simulations we employ the semi-impulsive limit where the pulse electric fields are delta functions in time, i.e. $E_n(\mathbf{r},t)=\delta(t)e^{-i\tilde{\omega}_n+i\mathbf{k_n r}}$. In this limit, the integrals over $\tau_i$ in the measured signal, e.g. the term in Eq.~\ref{eq:s_rp}, can be easily evaluated and $\sttt$ is simply equal to the response function,
\begin{equation}
    R\ttt = \mathrm{Tr}\{\hat{\mu}\, \mathcal{G}(t_3)[ \hat{\mu}, \mathcal{G}(t_2)[ \hat{\mu}, \mathcal{G}(t_1)[\hat{\mu}, \hat{\rho}_0]]]\}.
    \label{eq:def-response-tot}
\end{equation}
Here, $\mu(t)$ is the transition dipole operator, $\theta(t)$ is the Heaviside step function, $\rho_0$ is the density operator of the initial ground state, and ${\mathcal{G}(t) [\cdot] = e^{-i\hat{H}t} [\cdot]e^{i\hat{H}t}}$ denotes the time evolution of the operator ` $\cdot$ ' over a time interval $t$ with the Hamiltonian $\hat{H}$. In Sec.~\ref{sec:sim2des} we will test our GMM approach on simulated data where $\sttt=R\ttt$, but when comparing with experimental spectra we must account for finite-width pulse envelopes and therefore must use the more general expressions for $S\ttt$ shown for the rephasing contribution in Eq.~\ref{eq:s_rp} and for all terms in SI Sec.~\ref{si:sig_eqs}. 

To express the 2DES spectra in terms of the spectral density $\jw$ one can employ the second-order truncation of the cumulant expansion of the energy gap correlation function.\cite{mukamelPrinciplesNonlinearOptical1995,choTwoDimensionalOpticalSpectroscopy2009,zuehlsdorffOpticalSpectraCondensed2019,wiethornCondonLimitCondensed2023} This truncation is exact in the limit of Gaussian energy gap fluctuations and has previously been shown to be accurate in a range of systems.\cite{kellyTwoDimensionalElectronicSpectroscopy2025} While higher-order cumulants can be incorporated to improve model accuracy\cite{allanTamingThirdOrder2024}, they can introduce other issues such as spurious negative features in the linear absorption spectra. Here, we therefore use the second-order form, as it provides a physical foundation while retaining the flexibility to learn details from the 2DES information provided to the model. In addition, the model we develop employs the Condon approximation, although it could potentially be generalized.\cite{wiethornCondonLimitCondensed2023} We focus on systems that are well-approximated by two electronic states, but the extension to many-state electronic systems is possible. Within these approximations, the response function is~\cite{mukamelPrinciplesNonlinearOptical1995}
\begin{align}
    R\ttt= \left(\frac{i}{\hbar}\right)^3\theta(t_1)\theta(t_2)\theta(t_3)\sum_{i=1}^4 \mathrm{Im}\left[R_i\ttt\right].
    \label{eq:response-cumulant}
\end{align}
which can be written in terms of the four pathways $R_i$, each of which corresponds to a different physical process~\cite{mukamelPrinciplesNonlinearOptical1995}
\begin{subequations}
\label{eq:R_i}
\begin{align}
    R_1\ttt&=e^{-i\weg (t_3+t_1)}e^{-g^*(t_3)-g(t_1)-f_+\ttt} \label{eq:R_1}\\
    R_2\ttt&=e^{-i\weg (t_3-t_1)}e^{-g^*(t_3)-g^*(t_1)+f_+^*\ttt} \label{eq:R_2} \\
    R_3\ttt&=e^{-i\weg (t_3-t_1)}e^{-g(t_3)-g^*(t_1)+f_-^*\ttt} \label{eq:R_3} \\
    R_4\ttt&=e^{-i\weg (t_3+t_1)}e^{-g(t_3)-g(t_1)-f_-\ttt} \label{eq:R_4} 
\end{align}
\end{subequations}
where $R_1$ and $R_4$ are nonrephasing pathways, $R_2$ and $R_3$ rephasing pathways, $R_1$ and $R_2$ stimulated emission (SE), and $R_3$ and $R_4$ ground state bleach (GSB). Here $\weg$ is the thermal average energy gap, $g(t)$ the line shape function, and 
\begin{subequations}
\label{eq:f_pm}
\begin{align}
    f_+\ttt &= g(t_2)-g(t_2+t_3)- \nonumber \\ &\quad g(t_1+t_2)+g(t_1+t_2+t_3) \\
    f_-\ttt &= g^*(t_2) - g^*(t_2+t_3) -  \nonumber \\ &\quad g(t_1 + t_2) + g(t_1 + t_2 + t_3).
\end{align}
\end{subequations}
The line shape function is obtained from the spectral density $\jw$~\cite{mukamelPrinciplesNonlinearOptical1995}
\begin{align}
    g(t)&=\int_0^\infty\dd\omega \; \frac{\jw}{\omega^2}\left\{\coth\left(\frac{\beta\hbar\omega}{2}\right)\left[1-\cos(\omega t)\right]- \right.\nonumber \\ &\quad i\left[\sin(\omega t)-\omega t\right]\biggr\}.
    \label{eq:gt}
\end{align}
Equations~\ref{eq:polarization}-\ref{eq:gt} provide the connection between the spectral density and the signal that we will use in our GMM to build a model of the spectral density from limited 2DES information. The spectral density encodes how the nuclear motions of the system, both that of the solute and solvent, couple to electronic excitations. In addition to providing insight into vibronic couplings, $\jw$ can be used to reconstruct important physical quantities such as the reorganization energy. It can also be used to calculate the 2DES and other optical properties such as the linear absorption and pump-probe spectra. For example the linear absorption spectrum can be obtained from
\begin{align}
\label{eq:lin_abs}
    \sigma(\omega)\propto\mathrm{Re}\int_0^\infty e^{-i\weg t-g(t)}e^{i\omega t}\dd t
\end{align}
and the reorganization energy can be calculated as
\begin{align}
\label{eq:reorg_energy}
    \lambda=\frac{1}{\pi\hbar}\int_0^\infty\frac{\jw}{\omega}\dd \omega
\end{align}
As discussed in Sec.~\ref{sec:methods-gmm}, we use a GMM framework to fit the optimal spectral density from limited 2DES information, which can then be used to provide additional spectra and to extrapolate 2DES forward and backward in $t_2$.

\begin{figure*}[t]
    \centering
    \includegraphics[width=1\linewidth]{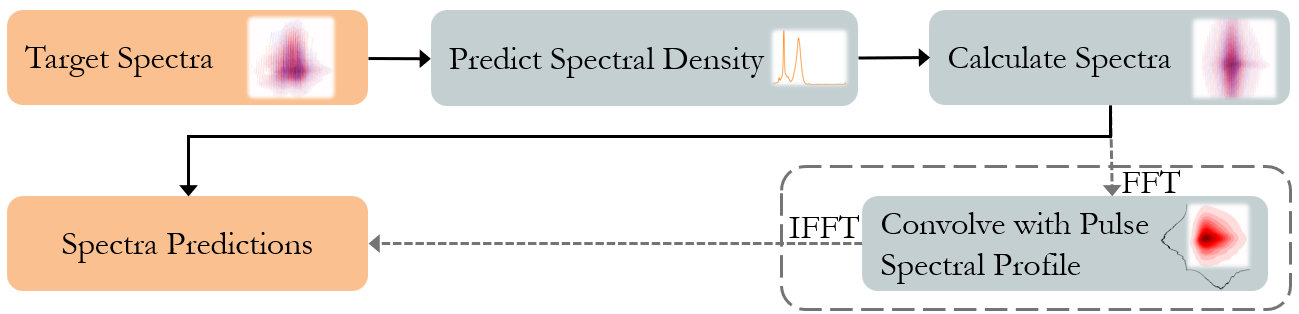}
    \caption{Our GMM framework. The GMM fits input target spectra in the $\sttw$ domain to model the system spectral density. The predicted spectral density is used to calculate 2DES and other spectra. If the GMM application is to experimental data, the pulse spectral profile is applied to capture its effect on the spectra.}
    \label{fig:gmm_schematic}
\end{figure*}

\subsection{Experimental Considerations in Modeling Electronic Spectroscopy}
\label{sec:methods-expt}

When considering simulated data, our framework only needs to utilize Eqs.~\ref{eq:R_i}-\ref{eq:gt} to calculate the predicted 2DES spectra as our simulated data employs the semi-impulsive limit. However, when comparing with experiments, the framework must account for convolution with the pulse electric fields and, where appropriate, the effects of phasing to ensure a faithful comparison between the reference and predicted spectra. To model the former, we take the experimental pulse spectral profiles $A_n(\omega)$ and use the convolution theorem to apply the pulse electric fields in frequency space by multiplying them by the predicted spectra $\swtw$ (SI Sec.~\ref{si:sig_eqs}).

In fully noncollinear experiments, such as the ones we apply our method to here (SI Sec.~\ref{si:nileblue-exp-det}), phase errors can originate from imperfect phase stability, challenges in achieving sub-wavelength accuracy of pulse timings, and small fluctuations at the short wavelengths of light used in 2DES.~\cite{brixnerPhasestabilizedTwodimensionalElectronic2004} In the pre-phased signal, $S^\mathrm{P}\wtw$, the real (absorptive) and imaginary (dispersive) parts are interwoven, making it difficult to disentangle the individual processes from the spectra and giving rise to erroneous features.~\cite{mengPostprocessingPhasecorrectionAlgorithm2017,brixnerPhasestabilizedTwodimensionalElectronic2004} To alleviate this issue, the signal can be phase-corrected to allow interpretation while maintaining separation between the absorptive and dispersive signal contributions. A common method for correcting the phase, which we use here, is through the projection-slice theorem~\cite{brixnerPhasestabilizedTwodimensionalElectronic2004,mengPostprocessingPhasecorrectionAlgorithm2017,singhIndependentPhasingRephasing2013}, which uses the proportional relationship between the pump-probe spectrum $I_\mathrm{PP}$ and the projection of the phase-corrected 2DES signal $\swtw$ onto the $\omega_3$ axis,
\begin{align}
    \label{eq:s_w3_proj}
    I_\mathrm{PP}(t_2,\omega_3)\propto\mathrm{Re}\int_{-\infty}^\infty \dd \omega_1 \swtw.
\end{align}
One therefore performs a subsequent pump-probe experiment to obtain the pump-probe spectrum $I_\mathrm{PP}(t_2,\omega_3)$. Equation~\ref{eq:s_w3_proj} can then be used to fit a phasing function that converts the pre-phased spectra $S^\mathrm{P}\wtw$ to its phase-corrected counterpart $\swtw$
\begin{align}
\label{eq:phase_fac}
I_\mathrm{PP}(t_2,\omega_3)&=N\;\mathrm{Re}\int_{-\infty}^\infty\dd \omega_1 \; S^\mathrm{P}\wtw \times \nonumber \\
&\quad e^{i\left[\phi_0+(\omega_3-\tilde{\omega}_3)\phi_1+(\omega_3-\tilde{\omega}_3)^2\phi_2^2\right]}
\end{align}
where $\tilde{\omega}_3$ is the center frequency of the $E_3$ pulse, and $N$ is a normalization constant.  The phase correction parameters, $\phi_0$, $\phi_1$, and $\phi_2$, are fit by minimizing the root-mean squared error loss between the directly-measured $I_\mathrm{PP}$ spectrum and the phase corrected projected 2DES signal~\cite{singhIndependentPhasingRephasing2013}. Physically, $\phi_0$ gives a phase constant shift, $\phi_1$ corrects for the timing uncertainty between pulse 3 and the local oscillator, and $\phi_2$ for phase distortions of the local oscillator.~\cite{singhIndependentPhasingRephasing2013,albrechtExperimentalDistinctionPhase1999}

\subsection{Gaussian Mixture Model Framework for 2DES Data}
\label{sec:methods-gmm}

GMMs provide a flexible framework for efficiently fitting high-dimensional data. Here we show how a GMM can be used to fit a model of the underlying spectral density of a system from a limited amount of 2DES information, which can then be used to predict 2DES and other optical spectra using the theory in Sec.~\ref {sec:methods-theory}, and also to provide an approach to predict which additional measurements will provide maximal information gain. Our GMM represents the spectral density as a sum of Gaussians, $\mathcal{G}(\mu,\sigma,a)$, parameterized by a mean $\mu$, variance $\sigma$, and amplitude $a$. These parameters are optimized on the optical spectroscopy (e.g. 2DES) data provided via minimization of a loss metric. Here, we predict and minimize on 2DES spectra in the $\sttw$ domain, for the reasons explained in Sec.~\ref{sec:results}.

To fit the GMM parameters for a given set of data we minimize a loss metric based on the Structural Similarity Index Measure (SSIM) between the current GMM prediction $y$ and the reference data $\hat{y}$,
\begin{align}
\label{eq:loss}
    \mathcal{L}_\mathrm{SSIM}(y, \hat{y}, c_1, c_2)&= \left< 1 - \frac{(2\mu_{\hat{y}}\mu_y + c_1)(2\sigma_{\hat{y}y} + c_2)}{(\mu_{\hat{y}}^2 + \mu_y^2 + c_1)(\sigma_{\hat{y}}^2 + \sigma_y^2 + c_2)}\right>
\end{align}
where $\mu_y = w * y$, $\sigma_y^2 = w * y^2 - \mu_y^2$, $\sigma_{\hat{y}y} = w * (\hat{y} \cdot y) - \mu_{\hat{y}}\mu_y$, $w$ is a Gaussian window, $c_1$ and $c_2$ are numerical stability constants, and $\langle\cdot\rangle$ takes the mean.

The loss is used to backpropagate and update the GMM parameters, and this process repeated iteratively until the loss is minimized. Using the GMM with the optimized parameters obtained from the minimization we then use the predicted spectral density to calculate the 2DES spectra as well as other electronic properties, such as the linear absorption spectrum and pump-probe spectra, in addition to calculating spectra at $t_2$ times other than the ones provided to assess how well the model is able to generalize to earlier and later time delays. 

Figure~\ref{fig:gmm_schematic} shows an outline of our GMM framework. The target 2DES spectra are provided to the GMM, which the model uses to fit the system's spectral density. The learned spectral density is then used to calculate 2DES spectra for any $t_2$ time delays of interest. If experimental factors are applied, the GMM then convolves the predicted spectra with the pulse spectral profile.

\subsection{Active Learning via Query by Committee}
\label{sec:query}

Query by committee (QbC) is an active learning strategy that uses an ensemble (committee) of models (members) that independently train on a given dataset to guide the selection of new data points. This is achieved by using the committee disagreement to guide which new data points will most improve model accuracy when added to the training set.~\cite{hinoActiveLearningQuery2023,freundInformationPredictionQuery1992} QbC has previously been used in various chemical fields, such as in training potential energy surfaces~\cite{smithLessMoreSampling2018,schranCommitteeNeuralNetwork2020} and materials discovery~\cite{hanInvDesFlowALActiveLearningbased2025,ringLeveragingDataMining2025}. QbC's utility is that it provides a proxy for the prediction error in cases where the ground truth may not be immediately available; e.g., for our application assessing the prediction accuracy for the 2DES signal at a given $t_2$ time delay requires performing a new experiment but by using QbC an estimate of the error at each $t_2$ can be made using only the standard deviation of the committee members. In practice, the procedure we employ involves:
\begin{enumerate}
    \item Using $M$ different random seeds to initialize GMMs from different initial spectral densities. SI Sec.~\ref{si:gmm_init} contains more details on our GMM initialization and training.
    \item Fitting GMM models to the same spectral dataset corresponding to a set of $t_2$ time delays resulting in $M$ GMMs (committee members).
    \item Calculate the time evolution ($t_2$) of the intensity at the \{$\omega_1,\omega_3$\} position of maximum intensity defined at $t_2$=0~fs, $\etmie{0}{t_2}$, for a range of $t_2$ time delays. \newchange{For the experimental data, we use the position of maximum intensity defined at $t_2$=80~fs, $\etmie{80}{t_2}$, since the experimental 2DES signal at earlier times contains significant contributions from the non-resonant response.}
    \item Calculate the standard deviation of the $\etmie{0}{t_2}$ predictions at each $t_2$.
    \item Select the $t_2$ for which the standard deviation is maximized to include into fitting.
\end{enumerate}

We use the committee disagreement in $\etmie{0}{t_2}$ to select the next $t_2$ time delay since it provides a measure of how the most dominant feature in the spectrum decays and how well this is captured by the current GMM prediction. To demonstrate the efficacy of this procedure, SI Fig.~\ref{fig:qbc} shows the improvement of 2DES prediction accuracy alongside the standard deviation (SD) of the $\etmie{0}{t_2}$ for the GFP in water, Nile red in benzene, and PYP in gas phase systems after training on their respective $t_2$=200~fs spectra. The $t_2$ time delays where $\etmie{0}{t_2}$ has the largest standard deviation is correlated with the $t_2$ time delays that improve the model accuracy most when added to the $t_2$=200~fs fitting, and hence these values were used.

\begin{figure}[h]
    \centering
    \includegraphics[width=1\linewidth]{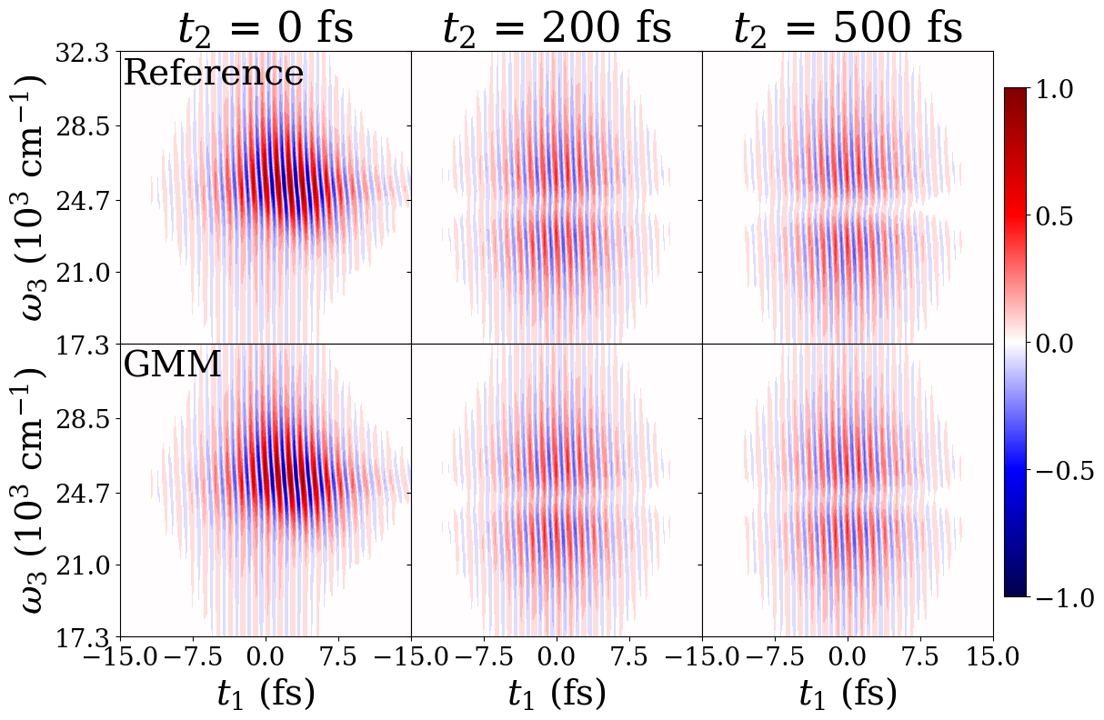}
    \caption{Comparison of the $\ttw$ spectra between the GMM prediction for the anionic GFP chromophore in water when fit only on $t_2$=200 fs. Top: The reference spectra. Bottom: The predicted spectra.}
    \label{fig:gfp_ttw}
\end{figure}

\section{Results and Discussion}
\label{sec:results}

In this section, we begin by assessing the performance of our GMM framework across several simulated systems and then apply it to experimental data. For the simulated systems, we show results for the anionic GFP chromophore in water, Nile red in benzene, and PYP in gas phase. These systems cover a wide range of conditions that might be encountered by chromophores in real chemical and biological systems, spanning those that interact strongly with their environments (GFP in water) to those that interact weakly (Nile red in benzene) or have no environment (PYP in gas phase). For each of these systems, we assess the accuracy of the framework in taking in the 2DES spectrum obtained either at a single time delay ($t_2$) or at multiple time delays. We use our framework to extrapolate the 2DES both forward and backward in time, predict the linear absorption and pump-probe spectra, and capture the system spectral density. We show that an active learning approach using QbC provides a strategy for picking additional $t_2$ measurements to include. We then assess how our framework generalizes to experimental systems by considering Nile blue in ethanol: we first apply experimental factors to demonstrate how these conditions impact the model performance, and then apply the framework to experimental spectra of Nile blue in ethanol.

\begin{figure*}[t!]
    \centering
    \includegraphics[width=1\linewidth]{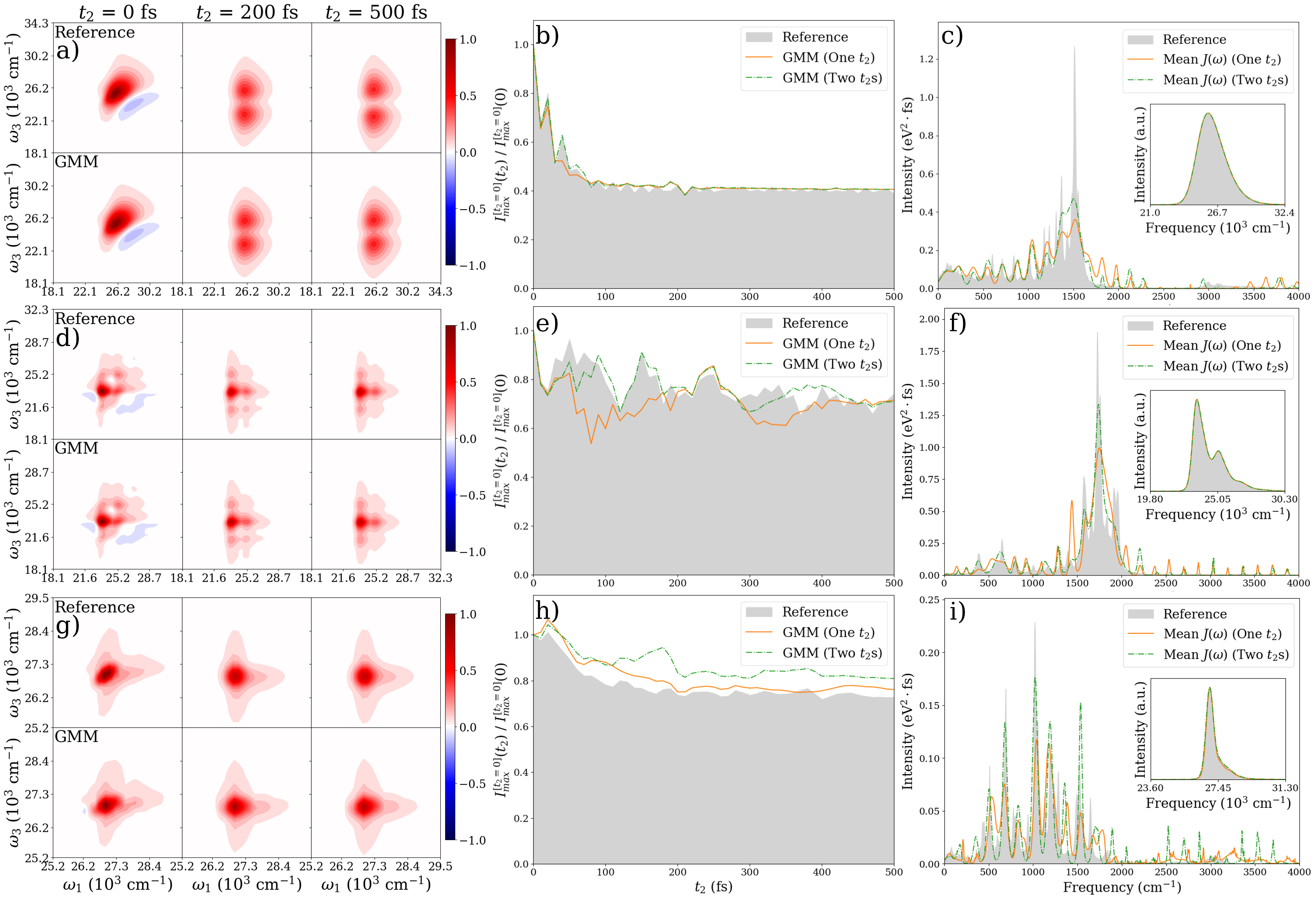}
    \caption{GMM predictions compared to the reference 2DES for Top: Anionic GFP chromophore in water, Middle: Nile red in benzene, Bottom: PYP in the gas phase. (a,d,g) GMM and reference 2DES spectra. (b,e,h) GMM and reference $\etmie{0}{t_2}$. (c,f,i) GMM and reference spectral densities and linear absorption spectra (inset).}
    \label{fig:sim_res}
\end{figure*}

\subsection{Application to Simulated 2DES}
\label{sec:sim2des}

We employ the GMM on simulated data of the anionic GFP chromophore in water~\cite{chenElucidatingRoleHydrogen2023,zuehlsdorffUnravelingElectronicAbsorption2018}, Nile red in benzene~\cite{chenExploitingMachineLearning2020}, and PYP in gas phase~\cite{zuehlsdorffUnravelingElectronicAbsorption2018} to demonstrate the flexibility of our framework in recapitulating the underlying spectral density and extrapolating the 2DES forward and backward in time using only minimal 2DES data. The anionic GFP chromophore in water has been shown to be a challenging test case for atomistic simulation methods~\cite{zuehlsdorffCombiningEnsembleFranckCondon2018,zuehlsdorffUnravelingElectronicAbsorption2018,chenElucidatingRoleHydrogen2023}, because it has strong interactions with the polar aqueous solvent, resulting in vibronic broadening of the system's spectra and strong coupling of solvent nuclear motions to electronic excitations. 

Figure~\ref{fig:gfp_ttw} shows the predicted $\sttw$ spectra for GFP in water at three different delay times when our GMM framework is fit to $\sttw$ at $t_2$=200~fs. \newchange{We selected $t_2$ = 200~fs as the initial training time delay for all our test systems, since this time is short enough that it is measured in most 2DES experiments while being long enough that it is well separated from experimental artifacts encountered around $t_2$ = 0~fs.} In principle, one could use the GMM approach described in Sec.~\ref{sec:methods-gmm} to fit a spectral density to the $\sttw$ or $\swtw$ signals. While the latter is more commonly shown, here we show results fit to $\sttw$ since experiments typically produce $\sttw$, which is then numerically Fourier transformed to $\swtw$, and hence, when only limited experiments have been performed, the Fourier transform across $t_1$ can introduce significant artifacts that one would want to avoid fitting. As shown in SI Sec.~\ref{si:wtw-fit}, for well-converged simulated data, fitting $\sttw$ data yields similar accuracy to fitting $\swtw$. Figure~\ref{fig:gfp_ttw} shows that our GMM accurately captures the reference spectrum it was fit to (i.e., the $t_2$=200~fs $\sttw$ spectrum) as well as the spectra at earlier ($t_2$=0~fs) and later ($t_2$=500~fs) time delays, which it was not directly fit to. 

We now consider the quality of the fits in the $\swtw$ domain, as this is the domain in which 2DES is typically analyzed. In addition, the $\sttw$ spectra are highly oscillatory in $t_1$, making it difficult to visually assess how closely the spectra align. Figure~\ref{fig:sim_res}(a) compares the reference and fit $\swtw$ spectra, where our framework accurately captures the spectra at time delays of $t_2$=0, 200, and 500~fs when trained only on the $t_2$=200~fs spectrum, i.e., both forward and backward in time. Thus, our method is able to accurately capture the sub-picosecond dynamics, which can often be the the most challenging to describe owing to the short timescales involved. The inset of Figure~\ref{fig:sim_res}(c) shows that the linear absorption is also captured \newchange{almost exactly} using a GMM fit using only $t_2$=200~fs 2DES information (orange line), even though it is not included in the fitting. Additionally, the GMM \newchange{reorganization energy (2190~$\cm$) accurately reproduced that computed from the reference (2201~$\cm$)}.

To provide a more detailed assessment of the 2DES time dependence, which is challenging to observe from just individual $t_2$ snapshots, Fig.~\ref{fig:sim_res}(b) shows the  $\etmie{0}{t_2}$, defined in Sec.~\ref{sec:query} as the time evolution ($t_2$) of the intensity at the \{$\omega_1,\omega_3$\} position of maximum intensity at $t_2$=0~fs. From this, one observes that, using a GMM fit to only $t_2$=200~fs (orange line), our framework accurately predicts the time dependence of $\etmie{0}{t_2}$ at short times and at longer time delays from $\sim$200~fs onward, but \newchange{exhibits minor discrepancies with the dynamics at intermediate times ($\sim$40~fs).} The spectral density, $\jw$, also shows some discrepancies when only a single $t_2$ is used, including broadened peaks and spurious high-frequency peaks (Fig.~\ref{fig:sim_res}(c), orange line). Hence, although it is hard to discern differences between the 2DES obtained from a GMM fit to a single $t_2$ time delay and the reference (Fig.~\ref{fig:sim_res}(a)), $\etmie{0}{t_2}$ and $\jw$ provide more strenuous tests. Our QbC approach (Sec.~\ref{sec:query}) selects $t_2$=40~fs as the additional time delay to include to improve the GMM (SI Fig~\ref{fig:qbc}) and indeed, the GMM incorporating this one extra time delay (green line) markedly improves the accuracy of the predicted $\etmie{0}{t_2}$ (Fig.~\ref{fig:sim_res}(a)) and $\jw$ (Fig.~\ref{fig:sim_res}(b)), with a slight decrease in the predicted reorganization energy to 2187~$\cm$. As shown in SI Sec.~\ref{si:expt2_fits}, including additional $t_2$ time delays leads to systematic convergence to the reference.

We now consider Nile red in benzene, which interacts more weakly with the solvent and thus has more evident vibronic progressions than the GFP system. Figure~\ref {fig:sim_res}(middle) shows our GMM performance when fit to a single $t_2$=200~fs 2DES spectrum of the simulated Nile red in benzene system, where our framework captures the $\swtw$ spectrum at $t_2$=200~fs and at earlier ($t_2$=0~fs) and later ($t_2$=500~fs) time delays. Likewise, it can infer the frequencies where the main $\jw$ intensity occurs (Fig.~\ref{fig:sim_res}(f), orange line), and predicts an accurate reorganization energy of 1254~$\cm$ (compared to the reference's 1243~$\cm$). Figure~\ref{fig:sim_res}(e), orange line, shows that the predicted $\etmie{0}{t_2}$ captures the relative intensities around $t_2$=200~fs but deviates at earlier and later time delays, although in this system it shows much less temporal decay than in GFP due to the weaker interactions between the chromophore and the solvent. In this case, using our QbC procedure suggests including the $t_2$ delay at 120~fs, which indeed results in our GMM better capturing the intensities both at earlier and later time delays (Fig.~\ref{fig:sim_res}(e), green line) while retaining an accurate reorganization energy of 1253~$\cm$.

The final simulated system we apply our GMM to is PYP in the gas phase, which, owing to the lack of solvent, has a narrow spectrum and structured spectral density. Figure~\ref{fig:sim_res}(g) shows our model can accurately capture the 2DES both forward and backward in time when fit to only $t_2$=200~fs. The model also captures the initial time decay of $\etmie{0}{t_2}$ (Fig.~\ref{fig:sim_res}(h), orange line), albeit plateauing at a slightly higher intensity than the reference. The spectral density, Fig.~\ref{fig:sim_res}(i), orange line, is well-described but has more spurious high-frequency peaks than seen in the other simulated examples. The predicted reorganization energy is 330~$\cm$, slightly lower than the reference's 351~$\cm$. Here, the QbC procedure indicates adding $t_2$=40~fs as an additional time delay to be fit, which results in improvement of the spectral density prediction, capturing the sharpness of the peaks and better predicting their intensities (Fig.~\ref{fig:sim_res}(i), green line), and reducing the error in the reorganization energy from -21~$\cm$ to +12~$\cm$. The additional $t_2$ delay gives slightly better structure to the predicted $\etmie{0}{t_2}$ but still fails to capture the intensity decay, being shifted upward with respect to the reference. 

Ultimately, when applied to simulated data, our GMM framework accurately reproduces spectroscopic observables, including the 2DES, linear absorption spectrum, and reorganization energy, across a wide range of solvation environments. Our method performs particularly well for condensed-phase systems, which are traditionally the most challenging yet the most relevant for a broad range of applications.

\begin{figure}[h]
    \centering
    \includegraphics[width=1\linewidth]{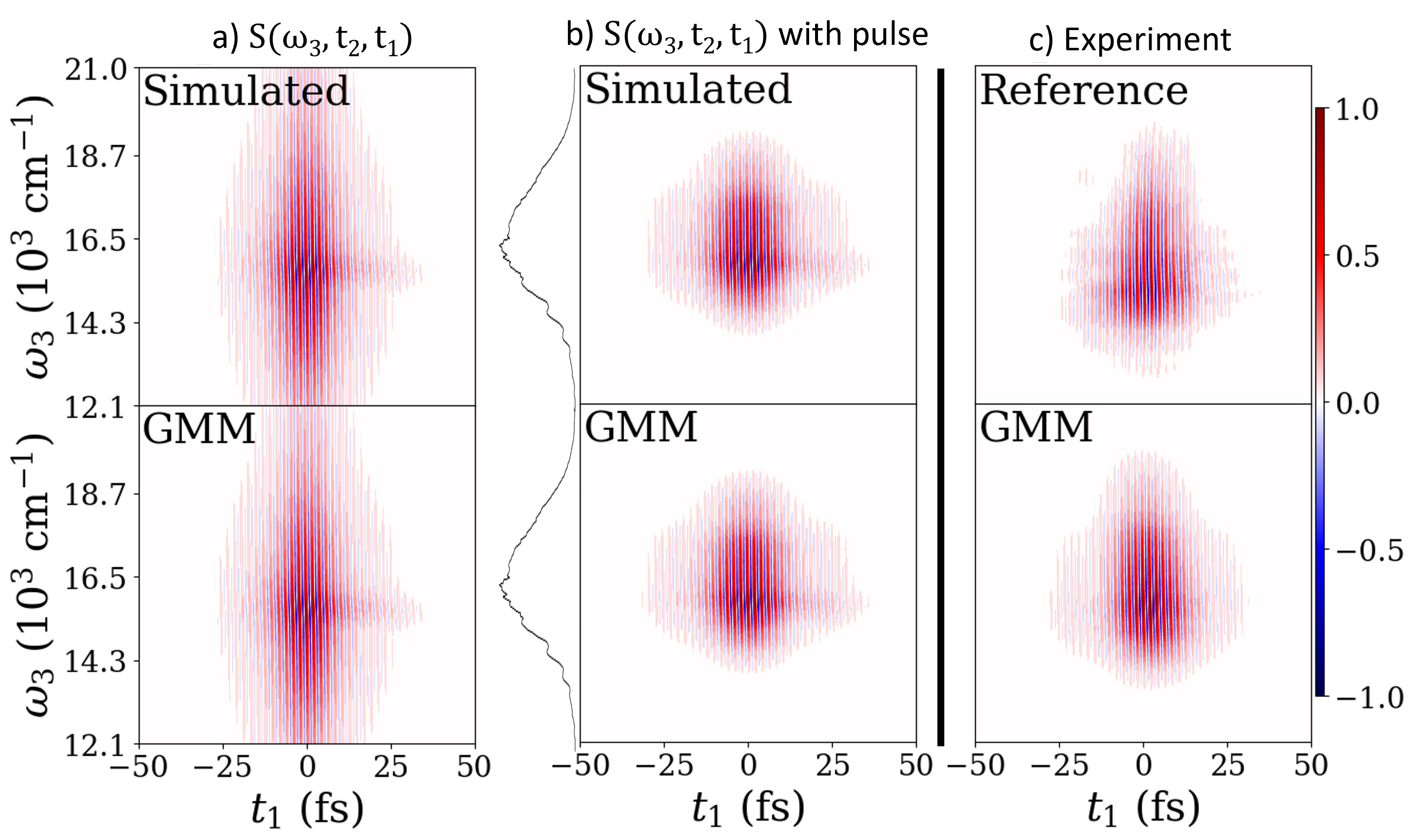}
    \caption{The effects of the pulse spectral profile and phasing on simulated 2DES of Nile blue in ethanol that are required for an accurate representation of the experiment, and the resulting GMM predictions. The top row shows the reference spectrum and the bottom row depicts the GMM prediction when fit to only the above spectrum. (a) Simulated 2DES of Nile blue in ethanol. (b) The simulated 2DES with the pulse spectral profile applied. (c) Experimental 2DES of Nile blue in ethanol with the pulse spectral profile applied to the GMM prediction.}
    \label{fig:nileblue_exp_considerations}
\end{figure}

\subsection{Application to Experimental 2DES}
\label{sec:exp2des}

\begin{figure*}[t]
    \centering
    \includegraphics[width=1\linewidth]{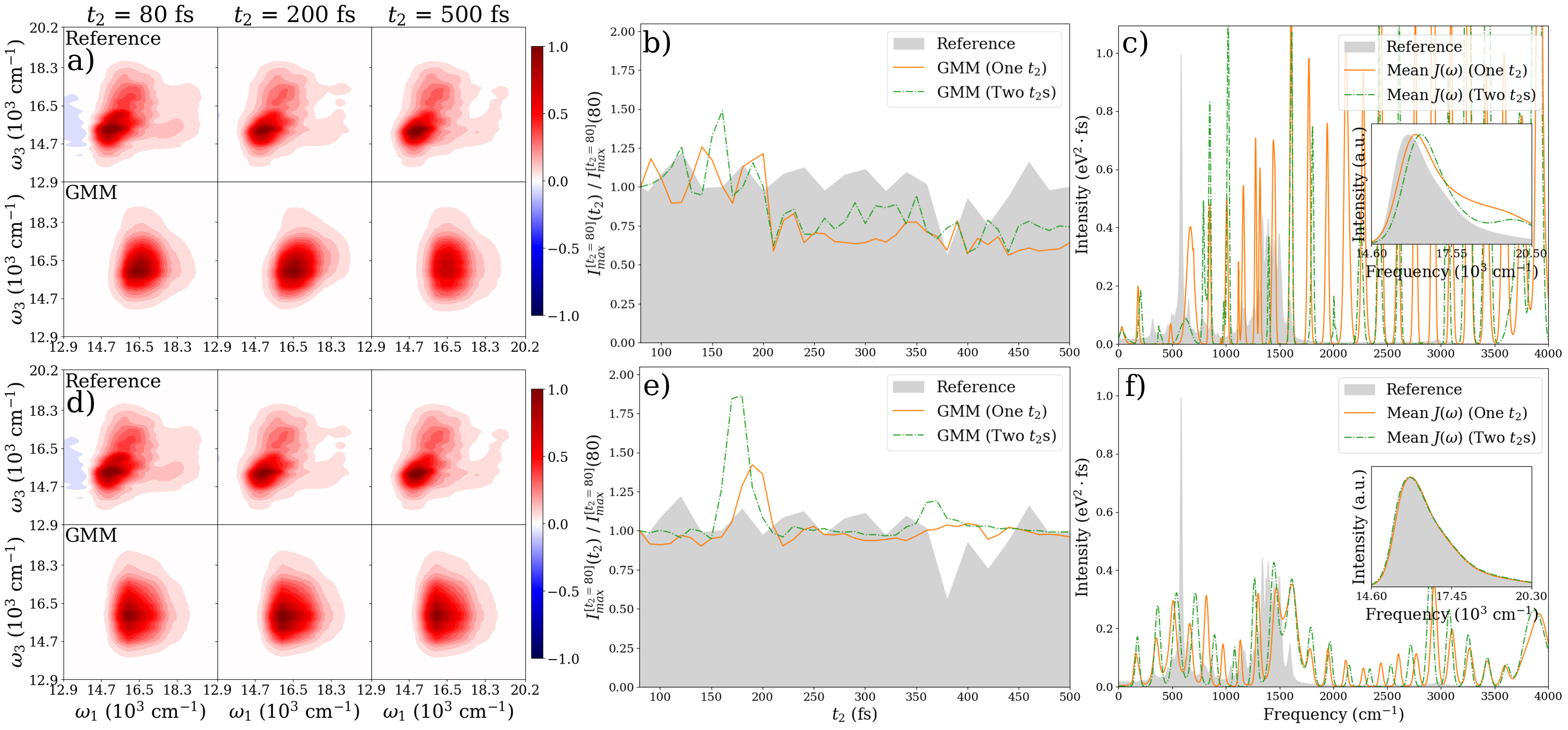}
    \caption{GMM predictions compared to the reference 2DES for experimental Nile blue in ethanol Top: fitting only the 2DES, Bottom: fitting the linear absorption alongside the 2DES. (a,d) GMM and reference 2DES spectra. (b,e) GMM and reference $\etmie{80}{t_2}$. (c,f) GMM and reference spectral densities and linear absorption spectra (inset). In panels c and f a simulated reference spectral density is shown as it is not experimentally measurable.}
    \label{fig:exp_res}
\end{figure*}

We now show how our GMM procedure can be extended to experimental systems and assess its performance on recent experiments performed for Nile blue in ethanol (SI Sec.~\ref{si:nileblue-exp-det}). Specifically, two major considerations that must be taken into account for the experimental data that are not present in the simulated data are: the finite width of the laser pulses and the phasing of the signal (Sec.~\ref{sec:methods-expt}).

To include the effects of the laser pulses in our fitting procedure, we multiply the GMM 2DES spectra prediction along $\omega_1$ and $\omega_3$ by the experimental pulse spectral profile (Sec.~\ref{sec:methods-theory}) before computing the loss function between the GMM prediction and the experiment. By doing this, the GMM prediction is not assessed in frequency ranges where experimental data is not measured. The application of the pulse spectral profile to a simulated $\sttw$ spectrum can be seen in Fig.~\ref{fig:nileblue_exp_considerations}(b), where its most pronounced effect is the narrowing of the signal in $\omega_3$.

The phasing of the 2DES signal is the second major consideration. There are various methods for phase-correcting 2DES~\cite{annaTwoDimensionalElectronicSpectroscopy2012,zhuCorrectionSpectralDistortion2019,lloydLeveragingScatterTwodimensional2020}, and we use the phase-corrected spectra as inputs. This allows experiments to be phase corrected using a user's method of choice prior to use with our framework. 

Applying our GMM to experimental Nile blue in ethanol after having accounted for the phase correction and pulse spectral profile, one can see in Fig.~\ref{fig:exp_res}(a) that the predicted 2DES captures the diagonal elongation at $t_2$=200~fs but does less well at capturing the higher-frequency signal contributions at approximately 16,000 - 18,500~$\cm$. The GMM is also able to capture some structure of the $\etmie{80}{t_2}$, although it shows a decrease in intensity at 200~fs that is not present in the reference. The spectral density is challenging to measure in experiments, particularly due to the fast decay of the frequency fluctuation correlation function limiting the spectral resolution, and hence in Fig.~\ref{fig:exp_res}(c) the reference shown is one obtained from recent simulations~\cite{kellyTwoDimensionalElectronicSpectroscopy2025}. However, it is clear that the GMM-predicted spectral density contains unphysical high-frequency peaks where no known vibrational modes are present in Nile blue or ethanol. This results in a larger reorganization energy (2310~$\cm$) than observed in experiment ($\sim$1010~$\cm$).~\cite{kellyTwoDimensionalElectronicSpectroscopy2025} These high frequencies in the GMM predicted spectral density also give rise to a linear absorption spectrum (Fig.~\ref{fig:exp_res}(c) inset) that is broader than observed experimentally. This is likely since the GMM is trying to fit features at high $\omega_{1}$ in the 2DES experiment that arise from residual phase errors, excited state absorption, and non-Condon fluctuations that are not included in the theory used here~\cite{kellyTwoDimensionalElectronicSpectroscopy2025,wiethornCondonLimitCondensed2023} by introducing high-frequency modes in the spectral density. Inclusion of an additional time delay using the QbC procedure at $t_2$=160~fs improves the predicted fluctuations within the $\etmie{80}{t_2}$ and narrows the predicted linear absorption spectrum, but does little to improve the strong, spurious peaks of the spectral density.

Since our GMM can predict spurious broadening of the linear spectrum at high frequencies and linear absorption spectra are routinely available and easy to measure, one might imagine using the latter to improve the former's predictive ability.
The bottom row of Fig.~\ref{fig:exp_res} shows the results obtained by including the linear absorption spectrum as an additional constraint when fitting the model by including it in the loss function (SI Sec.~\ref{si:lin_and_2d_fit}). Doing this leads to quantitative agreement with the linear absorption spectrum and reduces the erroneous high-frequency features in the spectral density, yielding a reorganization energy of 1254~$\cm$, which is much closer to the experimental value of 1010~$\cm$. However, the predicted 2DES (Fig.~\ref{fig:exp_res}(d)) fails to describe the diagonal elongation that was captured when the linear absorption spectrum was not included in the fit (Fig.~\ref{fig:exp_res}(a)), and it continues to fail to describe the higher-frequency contributions in $\omega_1$. The inclusion of an additional time delay, $t_2$=180~fs, selected via the QbC approach, in the training does little to improve the results. Hence, although the linear absorption spectrum provides an additional constraint, fitting the experimental 2DES at one or two time delays to an underlying spectral density remains challenging.

An additional constraint one could include is to weight the loss function based on the transient grating frequency-resolved optical gating (TG-FROG) spectrum, which provides a measure of the signal-to-noise ratio.~\cite{sonUltrabroadband2DElectronic2017,trebinoFrequencyResolvedOpticalGating2000} However, in SI Sec.~\ref{si:tgfrog-loss} we show that using the TG-FROG information to weight the loss function results in little improvement for Nile Blue in ethanol using our current GMM approach.

\begin{figure}[h]
    \centering
    \includegraphics[width=1\linewidth]{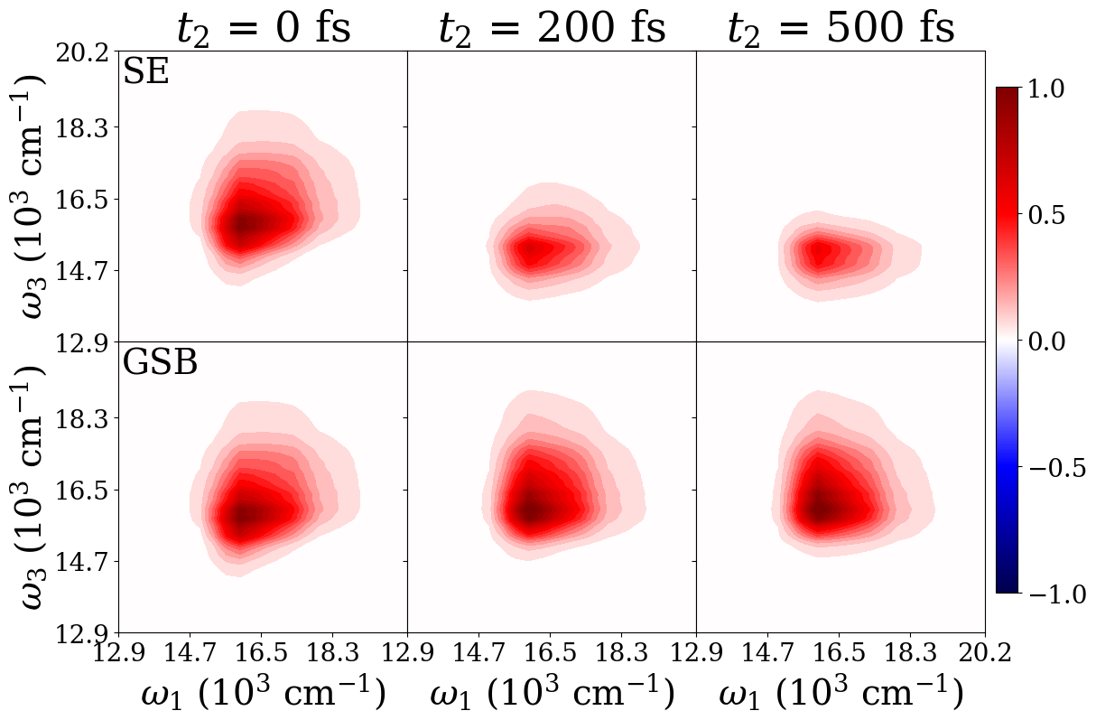}
    \caption{Decomposition of the GMM predictions for experimental Nile blue in ethanol with the linear absorption constraint into SE and GSB signal contributions.}
    \label{fig:se_gsb}
\end{figure}

In addition to our method providing access to spectroscopic observables, one can exploit the spectral density intermediate to elucidate signal contributions that are often difficult to separate experimentally. In particular, the 2DES signal is comprised of contributions from ground state bleach, stimulated emission, and excited state absorption effects. Hence, after learning the spectral density, one can exploit the second-order cumulant framework to separate these contributions. The decomposition of the GMM fit to the experimental data into the stimulated emission and ground state bleach contributions is shown in Fig.~\ref{fig:se_gsb}. This decomposition, along with the overall profile of the 2DES and excellent agreement with the linear absorption, demonstrate the power of our approach to streamline the simulation of these complex observables.

\section{Conclusion}
\label{sec:conclusion}

In this paper, we have shown that a Gaussian mixture model framework can be used to extract rich information from a small number of two-dimensional electronic spectroscopy measurements by learning the physics of the underlying system through its spectral density. We have shown that by using just a single 2DES time delay, our approach allows extrapolation of 2DES to earlier and later times and provides access to a wide range of other spectroscopic properties, such as the linear absorption spectrum and the reorganization energy. We have demonstrated the utility of this approach across a diverse range of chromophores spanning solvation environments that are strongly interacting (anionic GFP chromophore in water), weakly interacting (Nile red in benzene), and non-interacting (PYP in the gas phase), as well as on experimental data for Nile blue in ethanol. We have illustrated how our approach can be applied to experimental data by incorporating finite-width pulse-shape effects and phase correction into the model prediction, and have shown that fitting the linear absorption alongside 2DES can provide a physical constraint that dampens spurious high-frequency peaks in the spectral density that arise in the presence of noise. One could imagine extending this idea to include additional data that further constrains the fit, such as experimental pump-probe and fluorescence measurements, or data from atomistic ab initio simulations~\cite{kellyTwoDimensionalElectronicSpectroscopy2025}. Additionally, we have shown how an active-learning query-by-committee approach provides a data-efficient way to guide additional 2DES measurements by using GMM committee disagreement to identify which additional experiments are most likely to improve performance, without requiring access to additional data.

The GMM framework we have demonstrated here serves as a useful tool in predicting a system's spectral density, and while we have chosen to use this within the second-order cumulant framework for two-electronic-level systems, our model is generalizable beyond this. For example, extensions of our method to include additional electronic levels, higher-order cumulants, non-Condon effects, and theoretical frameworks beyond the cumulant expansion are possible. The spectral density learned by our GMM for a given system could also serve as an initial guess that could be refined using higher-level quantum dynamics methods, enabling a hierarchy of methods with increasing accuracy. Ultimately, we believe that this framework provides an efficient approach, requiring only 28 s on an H100 to provide a single fit, that can be used in tandem with experiments to elucidate and extrapolate from the data already collected and to suggest future experiments. \newchange{We have made our code and data, including 2DES spectra, GMM checkpoints, and scripts necessary to run with one's own data, available on GitHub\cite{MarklandGroupGmm2des} to facilitate future developments of machine learning-enhanced analysis and design of multidimensional spectroscopy experiments.}

\section*{Supplementary Material}

The supplementary material contains the following sections. S1: The equations that contribute to the measured 2DES signal. S2: The Nile blue in ethanol experimental details. S3: A description of how the GMM is initialized and trained. S4: A comparison between fitting the $\swtw$ spectra and the $\sttw$ spectra. S5: An assessment of our GMM performance when fitting simulated and experimental 2DES on 35 experimental $t_2$ delays. S6: How we fit the linear absorption spectrum alongside the 2DES. S7: The effect of weighting the SSIM loss by the TG-FROG spectrum on our GMM fit to experimental Nile blue in ethanol. S8: A description of how IR spectra were curated and processed to create a physically informed prior for our GMM Gaussian center initialization. S9: A description of our QbC procedure for selecting the next $t_2$ time delay. S10: An assessment of how the experimental pulse spectral profile affects our GMM fit to simulated 2DES. S11: Obtaining the thermal average energy gap from the linear absorption spectrum for a two-electronic-level system. \newchange{S12: Loss vs number of training steps for each of the systems discussed.}

\section*{Acknowledgments}

This work was funded by the National Science Foundation Grant No. CHE-2154291 to T.E.M. Joseph Kelly was also supported by a John Stauffer Memorial Award. This research used resources of the National Energy Research Scientific Computing Center (NERSC), a U.S. Department of Energy Office of Science User Facility located at Lawrence Berkeley National Laboratory, operated under Contract No. DE-AC02-05CH11231 using NERSC Award BES-ERCAP0035961. The experimental 2DES was supported by the US Department of Energy, Office of Science, Office of Basic Energy Sciences, Division of Chemical Sciences, Geosciences, and Biosciences under Award DE-SC0018097 to G.S.S.-C. A.M.C.~was supported by an Early Career Award in CPIMS program in the Chemical Sciences, Geosciences, and Biosciences Division of the Office of Basic Energy Sciences of the U.S.~Department of Energy under Award DE-SC0024154. M.S.C. was supported as a fellow of the Simons Center for Computational Physical Chemistry at NYU (SCCPC, Simons Foundation Grant MPS-T-MPS-00839534, MET).

\section*{Author declarations}

\subsection*{Conflict of interest}

The authors have no conflicts to declare.

\section*{Data availability}
\newchange{Code for training and evaluating the models can be found on GitHub at https://github.com/MarklandGroup/gmm2des. All other data that supports the findings of this study are available within the article and its supplementary material.}

\bibliography{references}
\newpage
\beginsupplement

\title{Supporting Information: Streamlining Analysis and Design of Two-Dimensional Electronic Spectroscopy using Machine Learning}

\author{Nicholas I. Hausman}
\affiliation{Department of Chemistry, Stanford University, Stanford, California, 94305, USA}

\author{Joseph Kelly}
\affiliation{Department of Chemistry, Stanford University, Stanford, California, 94305, USA}

\author{Michael S. Chen}
\affiliation{Simons Center for Computational Physical Chemistry, Department of Chemistry, New York University, New York, New York 10003, USA}

\author{Frank Hu}
\affiliation{Department of Chemistry, Stanford University, Stanford, California, 94305, USA}

\author{Angela Lee}
\affiliation{Department of Chemistry, Massachusetts Institute of Technology, Cambridge, Massachusetts 02139, USA}

\author{Andr\'es Montoya-Castillo}
\affiliation{Department of Chemistry, University of Colorado Boulder, Boulder, Colorado, 80309, USA}

\author{Gabriela S. Schlau-Cohen}
\email{gssc@mit.edu}
\affiliation{Department of Chemistry, Massachusetts Institute of Technology, Cambridge, Massachusetts 02139, USA}

\author{Thomas E. Markland}
\email{tmarkland@stanford.edu}
\affiliation{Department of Chemistry, Stanford University, Stanford, California, 94305, USA}

\maketitle
\tableofcontents

\section{Equations for the Contributions to the Signal}
\label{si:sig_eqs}
The signal in composed of a sum of eight contributions corresponding to different pathways.~\cite{mukamelPrinciplesNonlinearOptical1995,biswasCoherentTwoDimensionalBroadband2022,jonasTwoDimensionalFemtosecondSpectroscopy2003} These correspond to rephasing (RP), nonrephasing (NR), double quantum (DQ), and third-harmonic generation (THG) pathways, each of which can be measured by satisfying a particular phase-matching condition.
\begin{align}
\label{eq:si_sig}
    S_\mathrm{RP}\ttt&\propto e^{(-\mathbf{k}_1+\mathbf{k}_2+\mathbf{k}_3)\cdot \mathbf{r}} \int_0^\infty\dd\tau_3\int_0^\infty\dd\tau_2\int_0^\infty\dd\tau_1R\tttau\times \nonumber \\
    &\quad E_3(\mathbf{r},t_3-\tau_3)E_2(\mathbf{r},t_3+t_2-\tau_3-\tau_2)\times \nonumber \\
    &\quad E_1^*(\mathbf{r},t_3+t_1+t_2-\tau_3-\tau_2-\tau_1)\\
    S_\mathrm{NR}\ttt&\propto e^{(\mathbf{k}_1-\mathbf{k}_2+\mathbf{k}_3)\cdot \mathbf{r}} \int_0^\infty\dd\tau_3\int_0^\infty\dd\tau_2\int_0^\infty\dd\tau_1R\tttau\times \nonumber \\
    &\quad E_3(\mathbf{r},t_3-\tau_3)E_2^*(\mathbf{r},t_3+t_2-\tau_3-\tau_2)\times \nonumber \\
    &\quad E_1(\mathbf{r},t_3+t_1+t_2-\tau_3-\tau_2-\tau_1) \\
    S_\mathrm{DQ}\ttt&\propto e^{(\mathbf{k}_1+\mathbf{k}_2-\mathbf{k}_3)\cdot \mathbf{r}} \int_0^\infty\dd\tau_3\int_0^\infty\dd\tau_2\int_0^\infty\dd\tau_1R\tttau\times \nonumber \\
    &\quad E_3^*(\mathbf{r},t_3-\tau_3)E_2(\mathbf{r},t_3+t_2-\tau_3-\tau_2)\times \nonumber \\
    &\quad E_1(\mathbf{r},t_3+t_1+t_2-\tau_3-\tau_2-\tau_1) \\
    S_\mathrm{THG}\ttt&\propto e^{(\mathbf{k}_1+\mathbf{k}_2+\mathbf{k}_3)\cdot \mathbf{r}} \int_0^\infty\dd\tau_3\int_0^\infty\dd\tau_2\int_0^\infty\dd\tau_1R\tttau\times \nonumber \\
    &\quad E_3(\mathbf{r},t_3-\tau_3)E_2(\mathbf{r},t_3+t_2-\tau_3-\tau_2)\times \nonumber \\
    &\quad E_1(\mathbf{r},t_3+t_1+t_2-\tau_3-\tau_2-\tau_1)
\end{align}
Where $\mathbf{k}_i$ correspond to the wavevectors. The signal contributions can be isolated by satisfying the phase matching condition, where the signal is detected in the direction specified by the exponential (e.g., the RP signal is detected in the $\mathbf{k}=-\mathbf{k}_1+\mathbf{k}_2+\mathbf{k}_3$ direction).

\section{Nile Blue in Ethanol Experimental Details}
\label{si:nileblue-exp-det}
2DES measurements of Nile blue in ethanol were performed using a fully non-collinear, BOXCARS geometry setup. Full details can be found in \textit{Son, et. al.}\cite{sonUltrabroadband2DElectronic2017} In short, the 800 nm output out of a Ti:Saph regenerative amplifier (Coherent Libra) was passed through an argon tube (${\sim}$20 UNIT) wherein it undergoes self-phase modulation to generate white light. This white light is then compressed using double-angle chirp mirrors (Ultrafast Innovations) to a pulse width of ${\sim}$12 fs and usable spectral range of ${\sim}$550-700 nm, as determined by transient grating frequency-resolved optical
gating (TG-FROG). 

Nile blue was dissolved in ethanol to achieve an OD of ${\sim}$0.35 with a 0.2 mm path length cuvette (Starna), corresponding to a concentration of approximately ${\sim}$4.5 $\times$ 10$^{-4}$ M . 

2D spectra were taken at coherence times $\tau = -80-80$ fs in 0.4 fs steps and at waiting times $T = 0-100 $ fs in 6.67 fs steps, $T = 100-500$ fs in 20 fs steps, and $T = 500-1000$ fs in 50 fs steps. A negative $T$ time point was also taken ($T = -200$ fs) to determine the background component of the 2DES signal. The absolute value 2D spectra were phased using projection slice theorem \cite{jonasTwoDimensionalFemtosecondSpectroscopy2003}.

\section{Gaussian Mixture Model Initialization and Training}
\label{si:gmm_init}
\subsection{Initialization}
Our training procedure provides a Gaussian mixture model (GMM) that seeks to predict a spectral density consistent with the 2DES and/or the linear absorption spectrum. The spectral density was predicted over the range [0,~4000]~$\cm$ every 1~$\cm$ normalized in the code to span [0,~1], while the 2DES was fit over the range [0,~600]~fs every 1~fs for the \newchange{$t_1$ and $t_3$ time axes. The $t_2$ time delays fit were those specified in the main text.} The $t_3$ time axis was Fourier transformed to $\omega_3$ frequencies, resulting in predicted 2DES being on a $t_1\times\omega_3$ grid. The training data was interpolated to match the this grid to ensure time/frequency alignment during fitting.

\begin{figure}[H]
    \centering
    \includegraphics[width=1\linewidth]{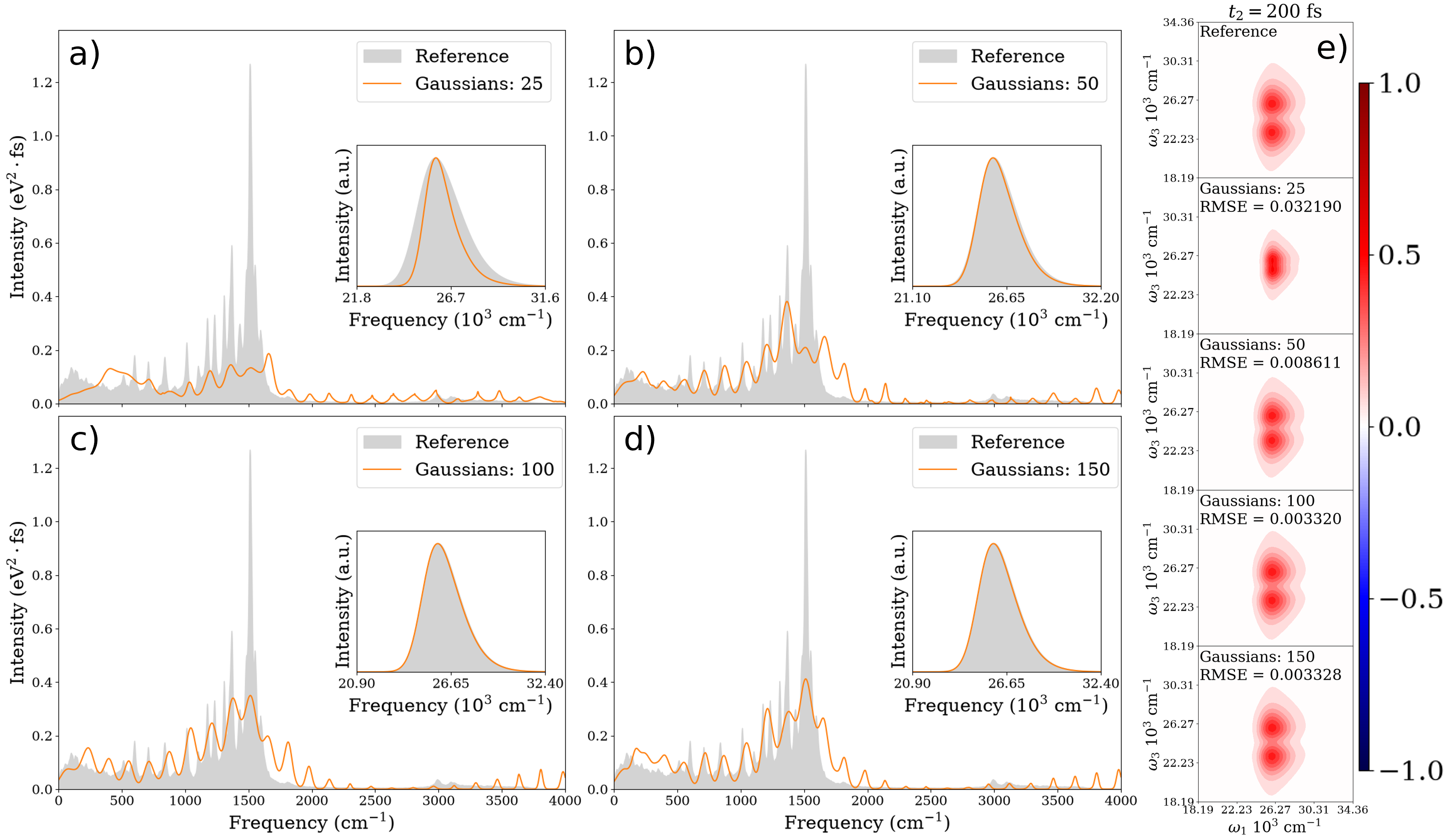}
    \caption{\newchange{Spectral density and linear absorption spectrum predictions for GFP in water fitting $t_2$=200~fs using (a) 25, (b) 50, (c) 100, and (d) 150 Gaussians. Panel (e) shows the corresponding 2DES predictions. Including more Gaussians improves the accuracy of the predicted 2DES up to $\sim$100 Gaussians.}}
    \label{fig:gaussian_scan}
\end{figure}

Our GMM has 301 parameters. These parameters consist of three for each of 100 Gaussians (the means $\mu_i$, variances $\sigma_i$, and amplitudes $a_i$) and 1 additional parameter, which is the average thermal energy gap $\weg$ of the system. \newchange{We found 100 Gaussians to be optimal for the systems studied. Figure~\ref{fig:gaussian_scan} shows fits to GFP in water when using 25, 50, 100, and 150 Gaussians, demonstrating the RMSE for the 100 Gaussian fit is the smallest.}
\alttext{Comparison between reference and GMM-predicted spectral densities for anionic GFP in water with a different number of Gaussians parameters, with corresponding 2DES predictions shown to the right. The GMM-predicted spectral density improves when including more Gaussian parameters up to 100 Gaussians, and more Gaussian parameters have diminished effect.}

To initialize the Gaussian parameters, the means were sampled from a distribution of typical vibrational frequencies observed in IR spectra, which was created by averaging a wide range of experimental IR spectra as detailed in SI Sec .~\ref {si:ir_spec}. The sampled means span a frequency range [0,4000]~$\cm$ and were then input into the GMM in the range [0,1] by dividing by the maximum frequency of 4000~$\cm$. The variances and amplitudes of the Gaussians were sampled from normal distributions $\mathcal{N}(10^{-2},10^{-6})$ and $\mathcal{N}(3\times10^{-2},10^{-6})$, respectively.

The average thermal energy gap was initialized by taking the weighted average position of the system's linear absorption spectrum, as this is analytically equivalent to the thermal energy gap for a system consisting of two electronic states (SI Sec.~\ref{si:wegav_from_linabs}).

\subsection{Training}
\begin{figure}[H]
    \centering
    \includegraphics[width=1\linewidth]{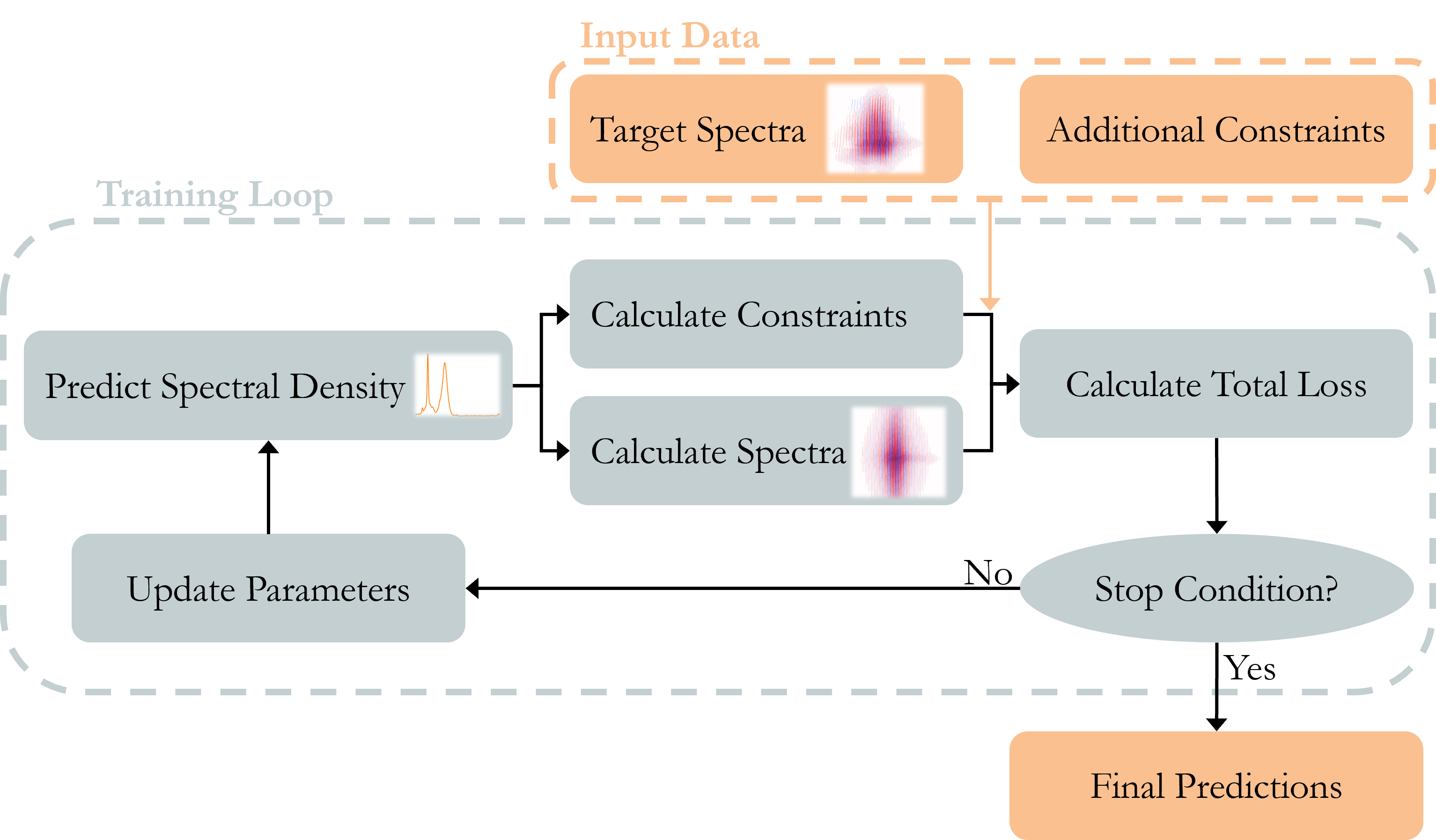}
    \caption{\newchange{Schematic depiction of the training loop for our GMM, starting with an initial spectral density prediction. The GMM uses the predicted spectral density to calculate 2DES along with additional constraints. Each prediction is used to calculate a loss with respect to the corresponding reference, and these are summed to get the total loss, which is used to update the GMM parameters. This repeats until a stopping condition is met (e.g., number of epochs, loss plateau), at which point the final predictions may be calculated from the final spectral density prediction.}}
    \label{fig:training_loop}
\end{figure}

Training of the GMM parameters were optimized using a structural similarity index measure (SSIM) loss metric (Eq.~\ref{eq:loss} in the main text) via gradient descent with an Adam optimizer and a learning rate of 0.001. Training was performed for 2000 epochs, which was found to be sufficient for observing plateauing the loss. For each system the results shown in the paper are those obtained form the lowest loss fit during the 2000 epochs. \newchange{Figure~\ref{fig:training_loop} provides a schematic of how our GMM is trained.}
\alttext{Flow diagram depicting the training loop for the GMM framework. The GMM predicts a spectral density which is used to calculate 2DES and other constraints. The predicted observables are compared with the target spectra and additional constraints to calculate a total loss. If a stop condition is met, the final predictions are made. Otherwise, the parameters are updated and the cycle repeats.}

\section{Fitting $\swtw$ Compared to Fitting $\sttw$}
\label{si:wtw-fit}
\begin{figure}[H]
    \centering
    \includegraphics[width=1\linewidth]{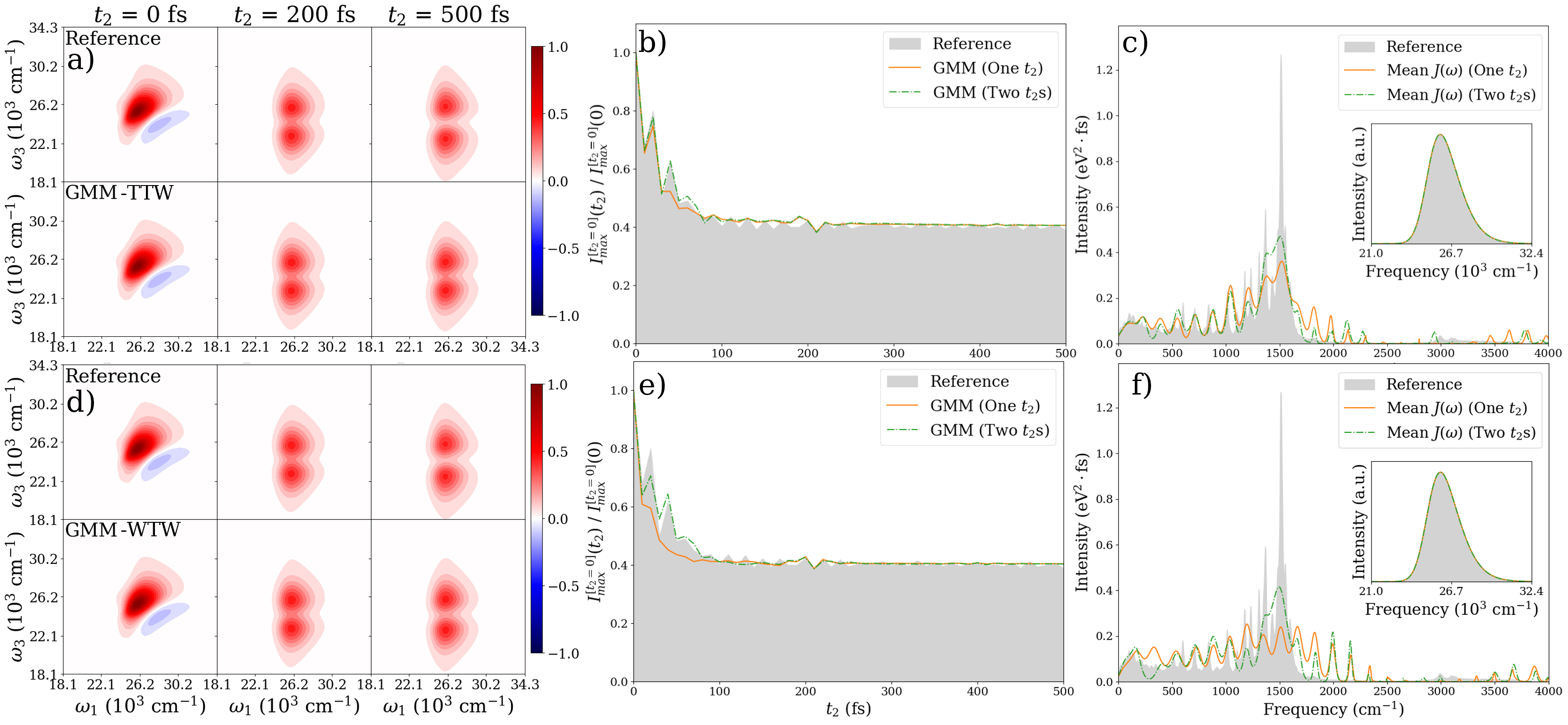}
    \caption{GMM predictions compared to the reference 2DES for anionic GFP chromophore in water. Top: Fitting the $\sttw$ spectra, Bottom: Fitting the $\swtw$ spectra. (a,d) GMM and reference 2DES spectra. (b,e) GMM and reference $\etmie{0}{t_2}$. (c,f) GMM and reference spectral densities and linear absorption spectra (inset).}
    \label{fig:gfp_wtw}
\end{figure}

\begin{figure}[H]
    \centering
    \includegraphics[width=1\linewidth]{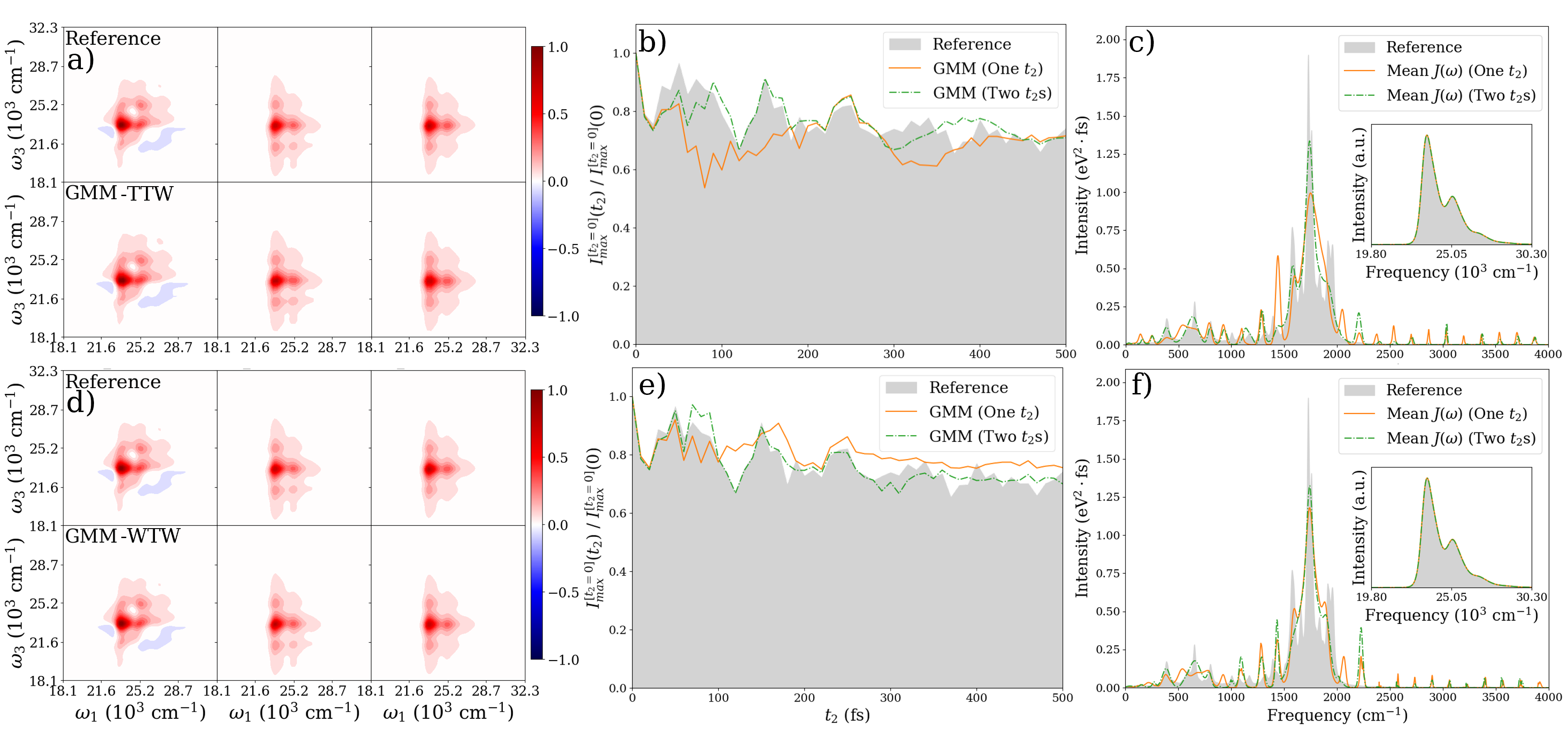}
    \caption{GMM predictions compared to the reference 2DES for Nile red in benzene. Top: Fitting the $\sttw$ spectra, Bottom: Fitting the $\swtw$ spectra. (a,d) GMM and reference 2DES spectra. (b,e) GMM and reference $\etmie{0}{t_2}$. (c,f) GMM and reference spectral densities and linear absorption spectra (inset).}
    \label{fig:nilered_wtw}
\end{figure}

\begin{figure}[H]
    \centering
    \includegraphics[width=1\linewidth]{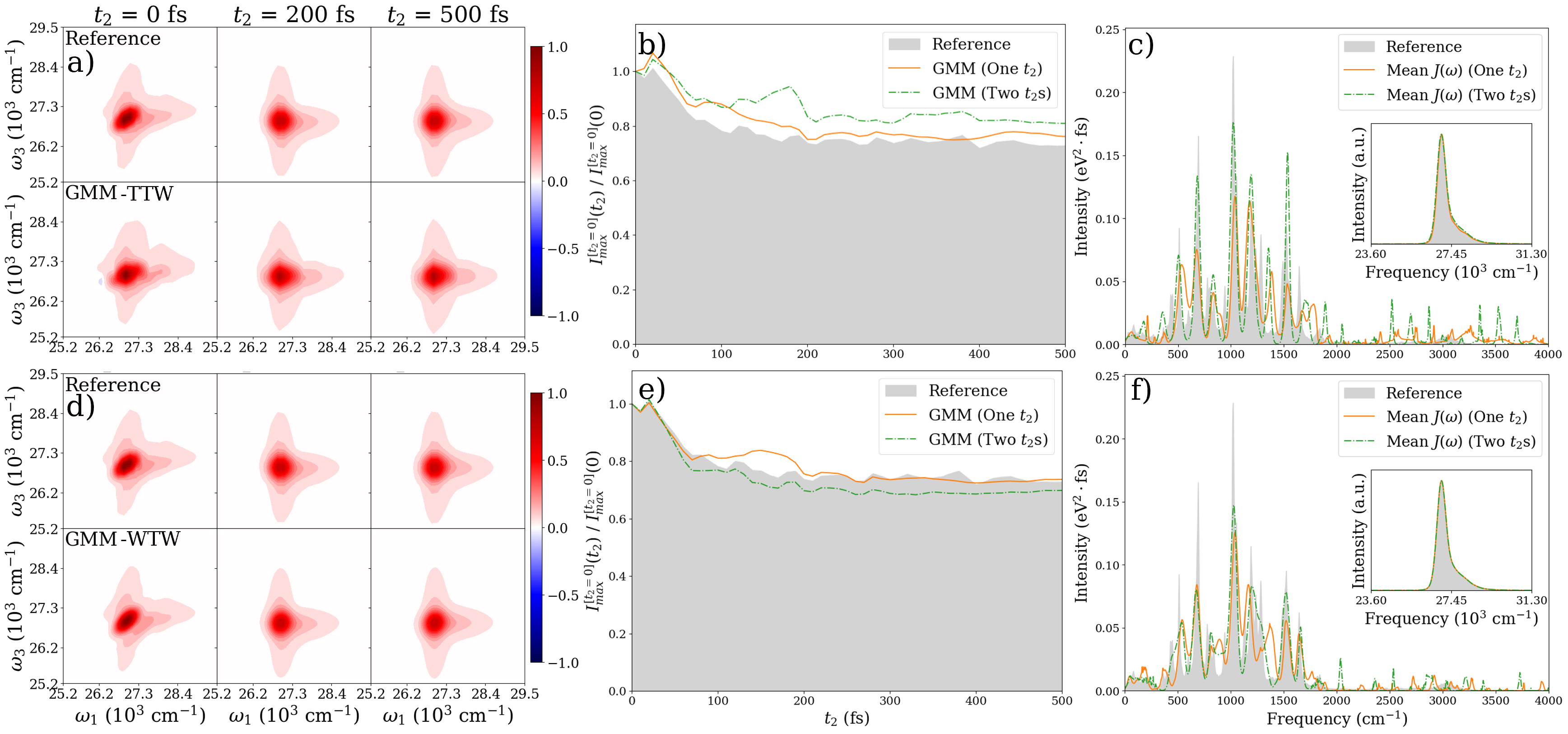}
    \caption{GMM predictions compared to the reference 2DES for PYP in the gas phase. Top: Fitting the $\sttw$ spectra, Bottom: Fitting the $\swtw$ spectra. (a,d) GMM and reference 2DES spectra. (b,e) GMM and reference $\etmie{0}{t_2}$. (c,f) GMM and reference spectral densities and linear absorption spectra (inset).}
    \label{fig:pyp_wtw}
\end{figure}

In the main text we show results when fitting $\sttw$ spectra. Here we show performance when instead fitting $\swtw$.
Fitting our GMM to $\swtw$ spectra gives similar accuracy to fitting the $\sttw$ spectra for simulated spectra of anionic GFP chromophore in water (Fig.~\ref{fig:gfp_wtw}), Nile red in benzene (Fig.~\ref{fig:nilered_wtw}), and PYP in gas phase (Fig.~\ref{fig:pyp_wtw}).
\alttext{Comparison of GMM and reference 2DES spectra, time-dependent maximum intensity, and spectral density with linear-absorption inset for anionic GFP in water, contrasting fits performed with interferogram (omega 3, t2, t1) spectra versus the frequency-frequency correlation maps (omega 3, t2, omega one). The two fitting domains give similar accuracy.}

\alttext{Comparison of GMM and reference 2DES spectra, time-dependent maximum intensity, and spectral density with linear-absorption inset for Nile red in benzene, contrasting fits performed with interferogram (omega 3, t2, t1) spectra versus the frequency-frequency correlation maps (omega 3, t2, omega one). The two fitting domains give similar accuracy.}

\alttext{Comparison of GMM and reference 2DES spectra, time-dependent maximum intensity, and spectral density with linear-absorption inset for PYP in gas phase, contrasting fits performed with interferogram (omega 3, t2, t1) spectra versus the frequency-frequency correlation maps (omega 3, t2, omega one). The two fitting domains give similar accuracy.}

\section{Fitting Experimental Population Times}
\label{si:expt2_fits}
In the main paper we show GMM performance fitting to either one or two $t_2$ time delays. Here we show the results of fitting the GMM to many (35) time delays. For the results shown below all the simulated and experimental systems were fit to the 35 time delays that are present in the experimental data: 80, 87, 93, 100, 120, 140, 160, 180, 200, 220, 240, 260, 280, 300, 320, 340, 360, 380, 400, 420, 440, 460, 480, 500, 550, 600, 650, 700, 750, 800, 850, 900, 950, 1000, 76808 fs. 

\begin{figure}[H]
    \centering
    \includegraphics[width=1\linewidth]{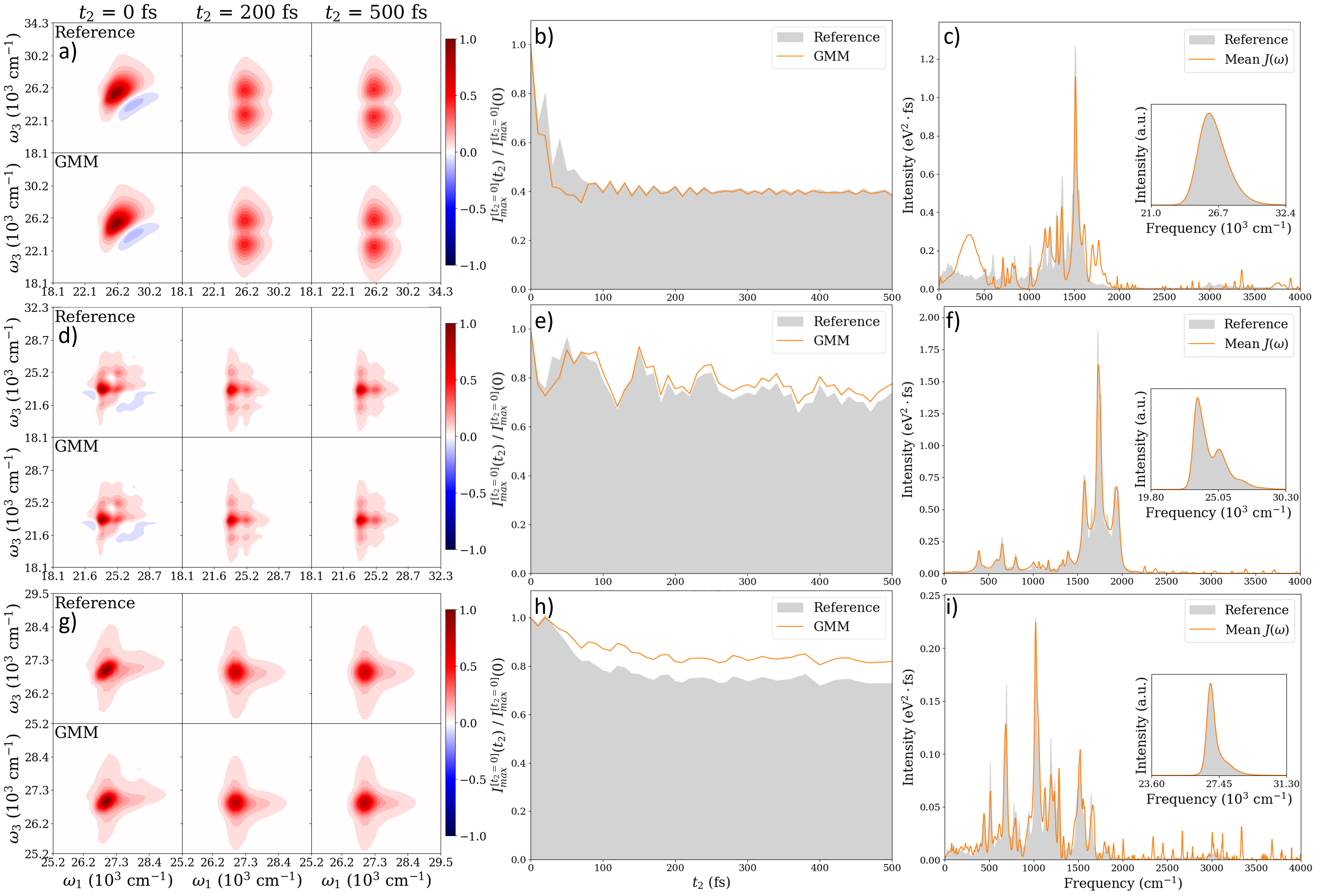}
    \caption{GMM predictions \newchange{for the simulated systems when fitting the same set of $t_2$ time delays as obtained in experiment} compared to the reference 2DES for Top: anionic GFP chromophore in water, Middle: Nile red in benzene, and Bottom: PYP in the gas phase. (a,d,g) GMM and reference 2DES spectra. (b,e,h) GMM and reference $\etmie{0}{t_2}$. (c,f,i) GMM and reference spectral densities and linear absorption spectra (inset).}
    \label{fig:sim_expt2s}
\end{figure}

For the simulated systems fit to the 35 time delays the results are shown in Figure~\ref{fig:sim_expt2s}. Figure~\ref{fig:sim_expt2s}{b} shows our GMM quantitatively captures the $\etmie{0}{t_2}$ from $\sim$100~fs onward when applied to anionic GFP chromophore in water. However, without access to $t_2$ time delays earlier than 80~fs, the model struggles to capture the earlier time fluctuations. The GMM is also able to describe the spectral density with great detail from approximately 600 - 1600~$\cm$ (Fig.~\ref{fig:sim_expt2s}(c)), but predicts overly intense low-frequency contributions. On both Nile red in benzene and PYP in the gas phase, our GMM was able to describe the $\etmie{0}{t_2}$, $\jw$, and linear absorption spectra with high fidelity, although the the predicted $\etmie{0}{t_2}$ for PYP does not decay to the proper relative intensity.

\alttext{GMM predictions compared to reference 2DES spectra, time-dependent maximum intensity, and spectral density with linear-absorption inset for anionic GFP in water, Nile red in benzene, and PYP in gas phase, each fit to all 35 available experimental time delays. The intensity and spectral density predictions match the references well, except for small discrepancies for anionic GFP in water, where the intensity is underpredicted prior to 100 femtoseconds and the low-frequency contributions to the spectral density are higher than the reference.}

\begin{figure}[H]
    \centering
    \includegraphics[width=1\linewidth]{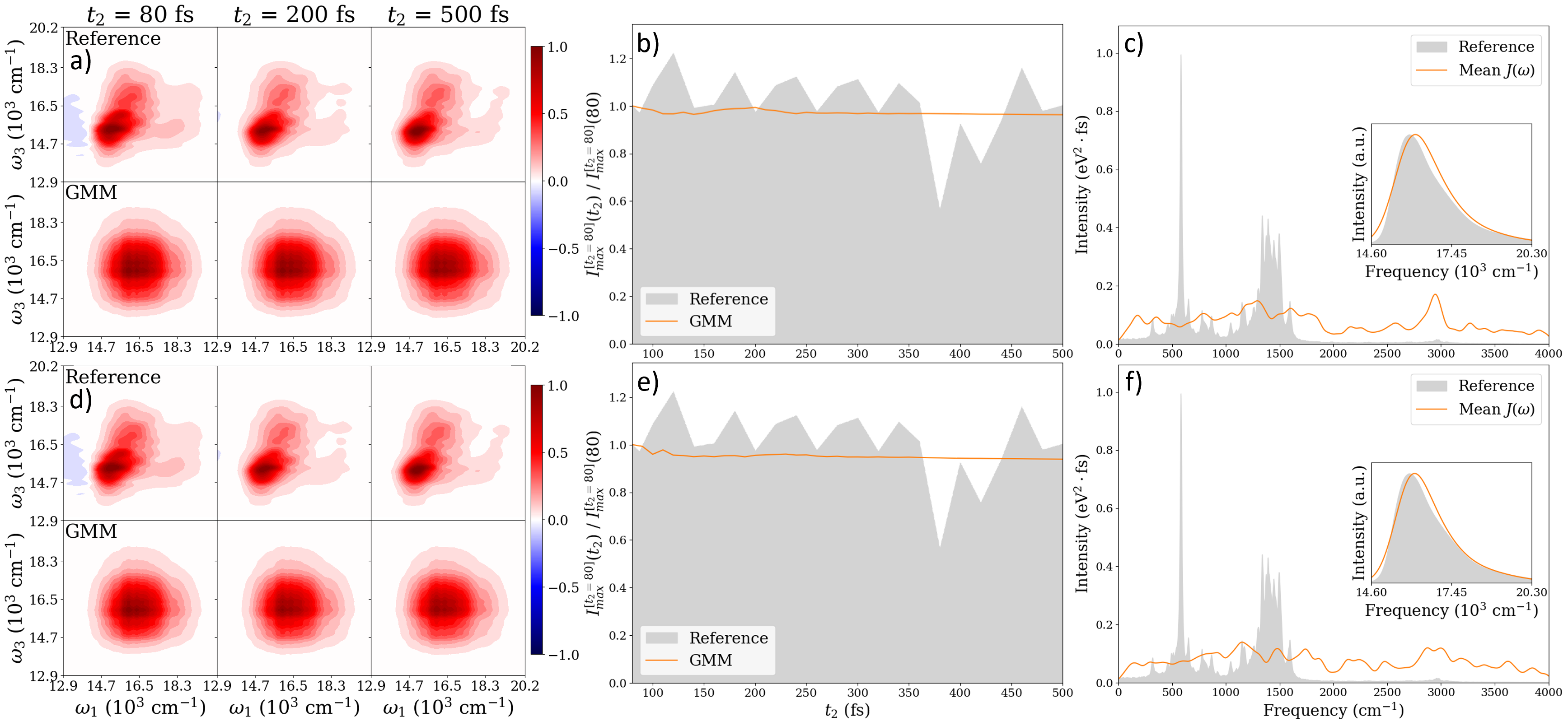}
    \caption{GMM predictions when fitting all the experimental $t_2$ time delays compared to the reference 2DES for experimental Nile blue in ethanol Top: fitting only the 2DES, Bottom: fitting the linear absorption alongside the 2DES. (a,d) GMM and reference 2DES spectra. (b,e) GMM and reference $\etmie{80}{t_2}$. (c,f) GMM and reference spectral densities and linear absorption spectra (inset). In panels c and f a simulated reference spectral density is shown as it is not experimentally measurable.}
    \label{fig:exp_expt2s}
\end{figure}

The results for the experimental system fit to 35 time delays are shown in Fig.~\ref{fig:exp_expt2s}. Our GMM predicts a globular 2DES (Fig.~\ref{fig:exp_expt2s}(a,d)) with little spectral evolution (Fig.~\ref{fig:exp_expt2s}(b,e)) regardless of the linear absorption constraint. Likewise, the model predicts a $\jw$ with moderate intensity at nearly all frequencies (Fig.~\ref{fig:exp_expt2s}(c,f)).

\alttext{2DES spectra, time-dependent maximum intensity, and spectral density with linear-absorption inset for experimental Nile blue in ethanol fit to all 35 experimental time delays, without (top) and with (bottom) the linear-absorption constraint. The GMM predicts a globular 2DES with little spectral evolution and a spectral density of moderate intensity across nearly all frequencies regardless of the constraint.}

\section{Fitting Linear Absorption Alongside 2DES}
\label{si:lin_and_2d_fit}
The linear absorption spectrum can be fit alongside the GMM 2DES predictions as an additional physical constraint on the system. We predict the linear absorption spectrum using Eq.~\ref{eq:lin_abs} in the main text and a root-mean-squared error (RMSE) loss metric. This is added to the SSIM loss over the 2DES (Eq.~\ref{eq:loss} in the main text) with a weighting term, which we chose to be $\alpha$=10:
\begin{align}
\label{sieq:lin_loss}
\mathcal{L}_{Total}=\mathcal{L}_{SSIM}(y,\hat{y})+\alpha\sqrt{\frac{1}{N}\sum_{i=1}^N (\sigma(\omega_i)-\hat{\sigma}(\omega_i))^2}.
\end{align}
Here, $y$ is the reference 2DES, $\hat{y}$ is the predicted 2DES, $\sigma(\omega)$ is the reference linear absorption spectrum, $\hat{\sigma}(\omega)$ is the predicted linear absorption spectrum, and the mean is taken over the $N$ frequency points.

\section{Weighting 2DES Loss by TG-FROG}
\label{si:tgfrog-loss}
In the main text we utilized the pulse spectral profile to mitigate loss being taken within the range of 2DES frequencies that were not detected by experimental measurement. One may be inclined to perform a similar transformation utilizing the full TG-FROG spectrum, as this spectrum is proportional to the signal-to-noise ratio. However, as shown in Figs.~\ref{fig:exp-res-tgfrog-lin} and \ref{fig:exp-res-tgfrog-nolin} below, we observed that weighting the $\mathcal{L}_{SSIM}$ metric by the TG-FROG spectrum resulted in minimal change when training with (Fig.~\ref{fig:exp-res-tgfrog-lin}) or without (Fig.~\ref{fig:exp-res-tgfrog-nolin}) the linear absorption constraint.

\begin{figure}[H]
    \centering
    \includegraphics[width=1\linewidth]{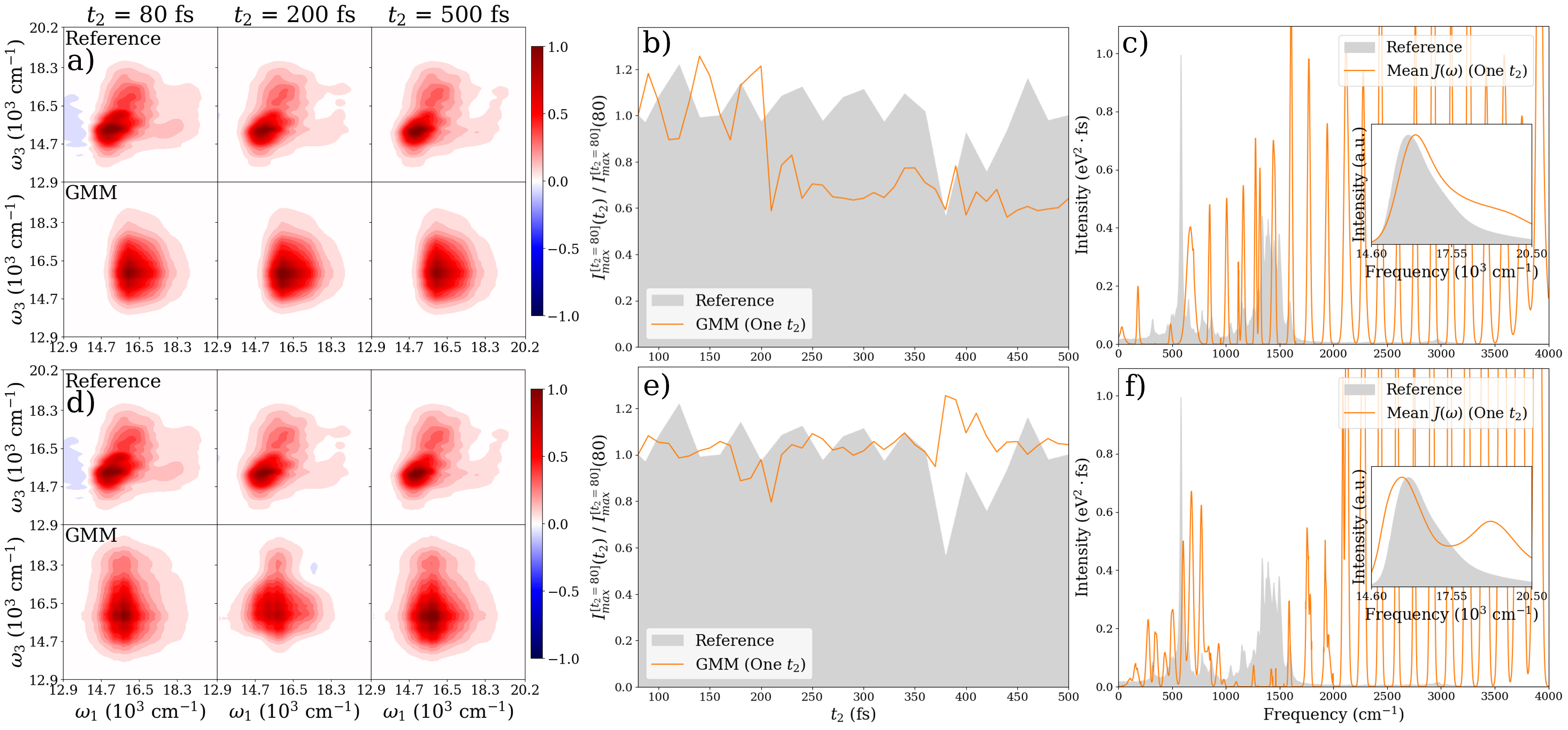}
    \caption{GMM predictions compared to the reference 2DES for experimental Nile blue in ethanol. Top: without weighting $\mathcal{L}_{SSIM}$ by the TG-FROG spectrum, Bottom: weighting $\mathcal{L}_{SSIM}$ by the TG-FROG spectrum. (a,d) GMM and reference 2DES spectra. (b,e) GMM and reference $\etmie{80}{t_2}$. (c,f) GMM and reference spectral densities and linear absorption spectra (inset). In panels c and f a simulated reference spectral density is shown as it is not experimentally measurable.}
    \label{fig:exp-res-tgfrog-nolin}
\end{figure}
\alttext{2DES spectra, time-dependent maximum intensity, and spectral density with linear-absorption inset for experimental Nile blue in ethanol without the linear-absorption constraint, contrasting fits with and without weighting the loss function by the TG-FROG spectrum. Weighting by the TG-FROG spectrum produces results similar to those in the main text.}

\begin{figure}[H]
    \centering
    \includegraphics[width=1\linewidth]{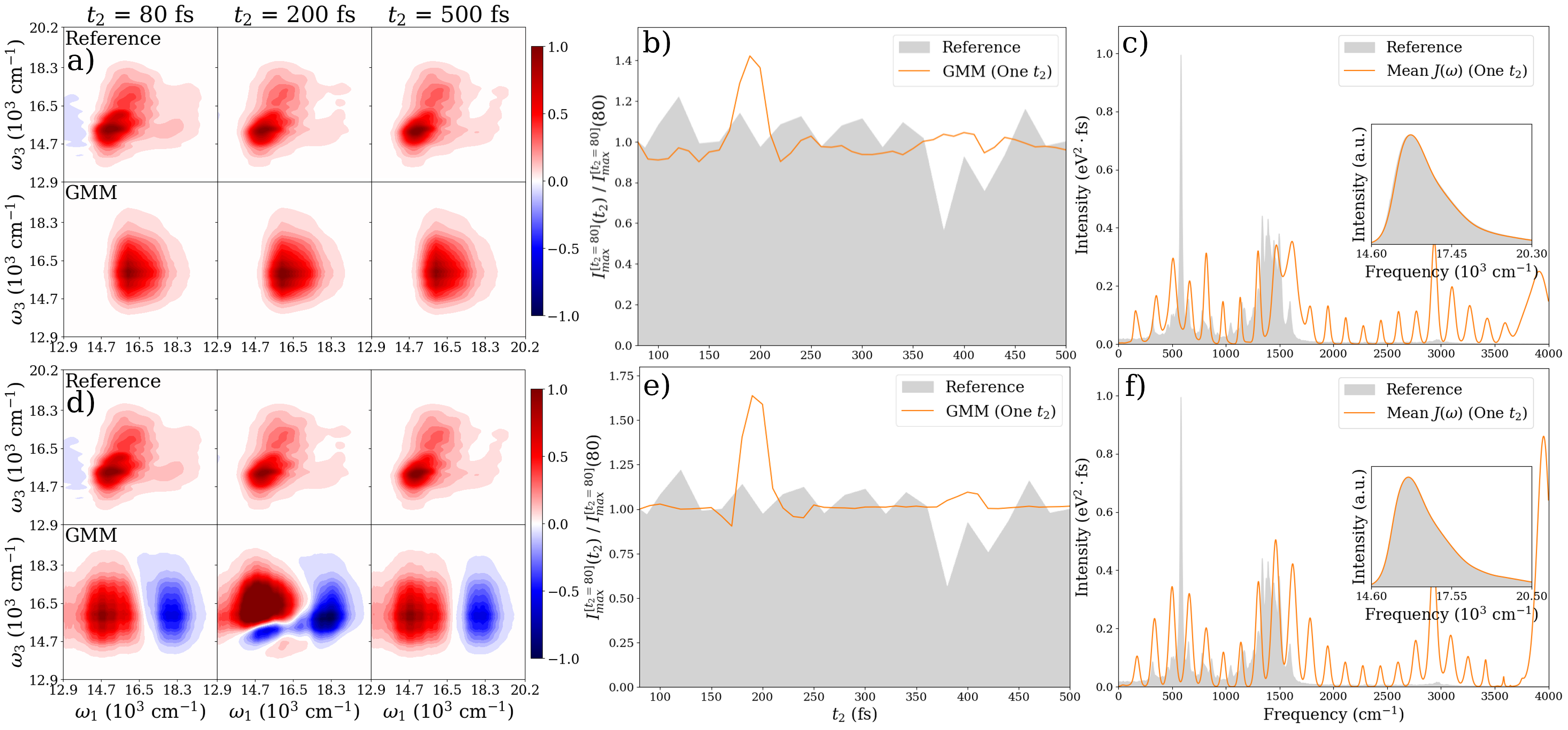}
    \caption{GMM predictions compared to the reference 2DES for experimental Nile blue in ethanol when fitting the linear absorption alongside the 2DES. Top: without weighting $\mathcal{L}_{SSIM}$ by the TG-FROG spectrum, Bottom: weighting $\mathcal{L}_{SSIM}$ by the TG-FROG spectrum. (a,d) GMM and reference 2DES spectra. (b,e) GMM and reference $\etmie{80}{t_2}$. (c,f) GMM and reference spectral densities and linear absorption spectra (inset). In panels c and f a simulated reference spectral density is shown as it is not experimentally measurable.}
    \label{fig:exp-res-tgfrog-lin}
\end{figure}
\alttext{2DES spectra, time-dependent maximum intensity, and spectral density with linear-absorption inset for experimental Nile blue in ethanol with the linear-absorption constraint included, contrasting fits with and without weighting the loss function by the TG-FROG spectrum. Weighting by the TG-FROG spectrum produces results similar to those in the main text.}

\section{IR Spectra as a Prior for Gaussian Centers}
\label{si:ir_spec}
To sample the GMM Gaussian means from a physically-motivated prior, we average experimental IR spectra to create a distribution of vibrational modes that is consistent with the positions of these frequencies in typical chemical systems.
The IR spectra used were obtained from the NIST Chemistry WebBook\cite{linstromNISTChemistryWebBook}. We gathered a total of 19,582 spectra which were processed in the following manner:
\begin{itemize}
    \item Only entries defined as gas, vapor, liquid, solution, melted, melt, oil, saturated, salted were kept. This reduced the number of spectra to 13749.
    \item Transmission spectra were converted to absorption spectra. Spectra with length mismatches between the wavenumbers and intensities were discarded. This reduced the number of spectra to 13702.
    \item Only those spectra with a maximum frequency greater than 2000~$\cm$ were included. This reduced the number of spectra to 12415.
    \item The spectra were interpolated to the range [0,~4000]~$\cm$ with a spacing of 1~$\cm$.
    \item Due to noise in the low frequency region ($<$ 100~$\cm$), each spectrum was multiplied by the hyperbolic tangent function $\tanh(\omega / 75~\cm)$, smoothing the low frequency region and enforcing zero signal at zero frequency.
    \item The spectra were averaged and divided by the sum of the average to create a probability distribution of vibrational modes.
\end{itemize}

\section{Determining Next Training Point using Committee Uncertainty}
\label{si:qbc}
We used a query-by-committee (QbC) active learning approach to determine the next 2DES time delay ($t_2$) to train our GMM on. To achieve this we trained an ensemble of $M$ GMM members on the same reference 2DES starting from different initializations (SI Sec.~\ref{si:gmm_init}). 

\begin{figure}[H]
    \centering
    \includegraphics[width=0.5\linewidth]{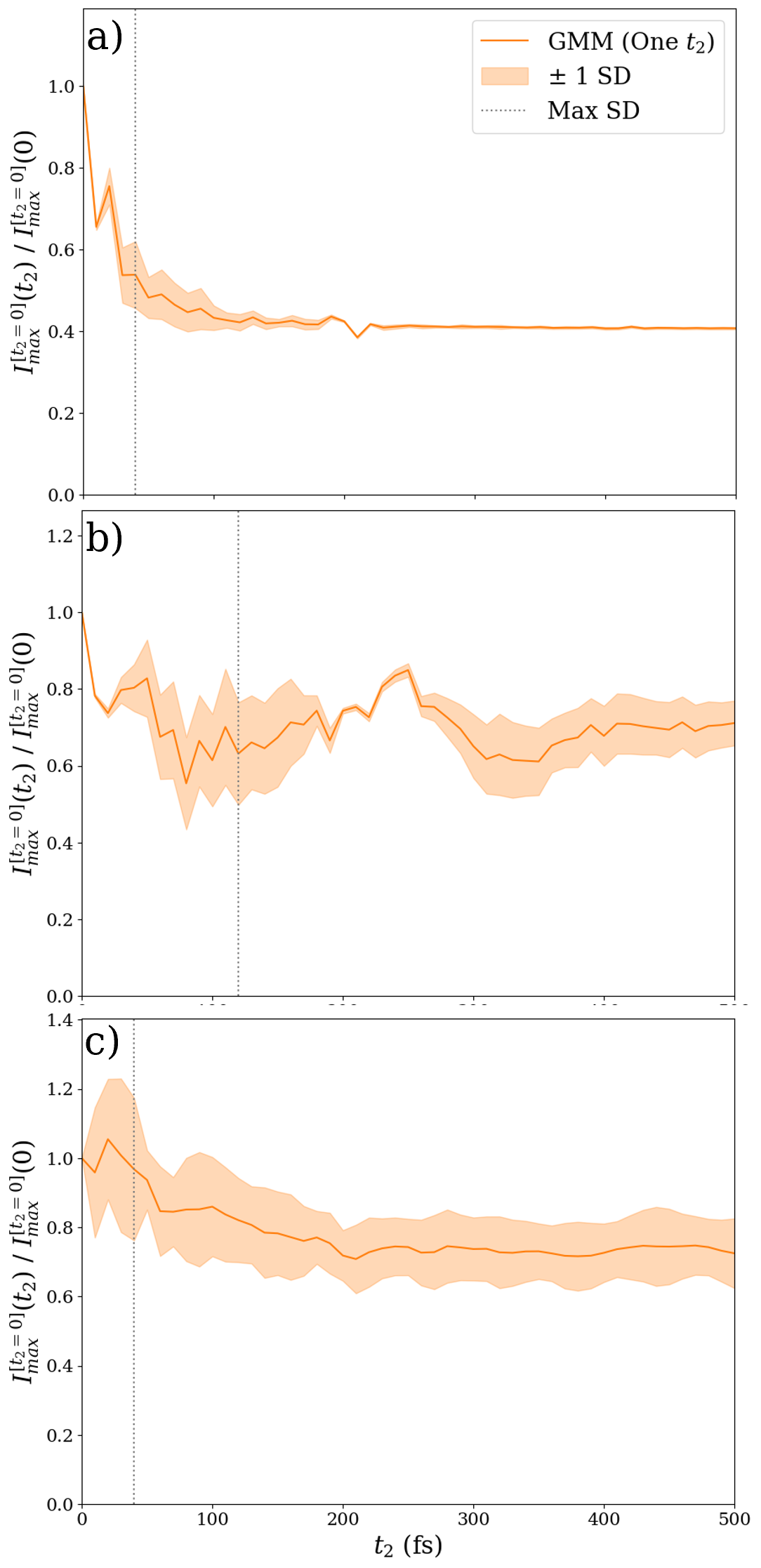}
    \caption{The mean predicted $\etmie{0}{t_2}$ of the M=10 GMM committee members with uncertainty shown as $\pm$1 SD in the predictions for (a) anionic GFP in water, (b) Nile red in benzene, and (c) PYP in the gas phase. The vertical grey dashed lines indicate the $t_2$ time delay of maximum uncertainty.}
    \label{fig:uncertain}
\end{figure}

\begin{figure}[H]
    \centering
    \includegraphics[width=0.5\linewidth]{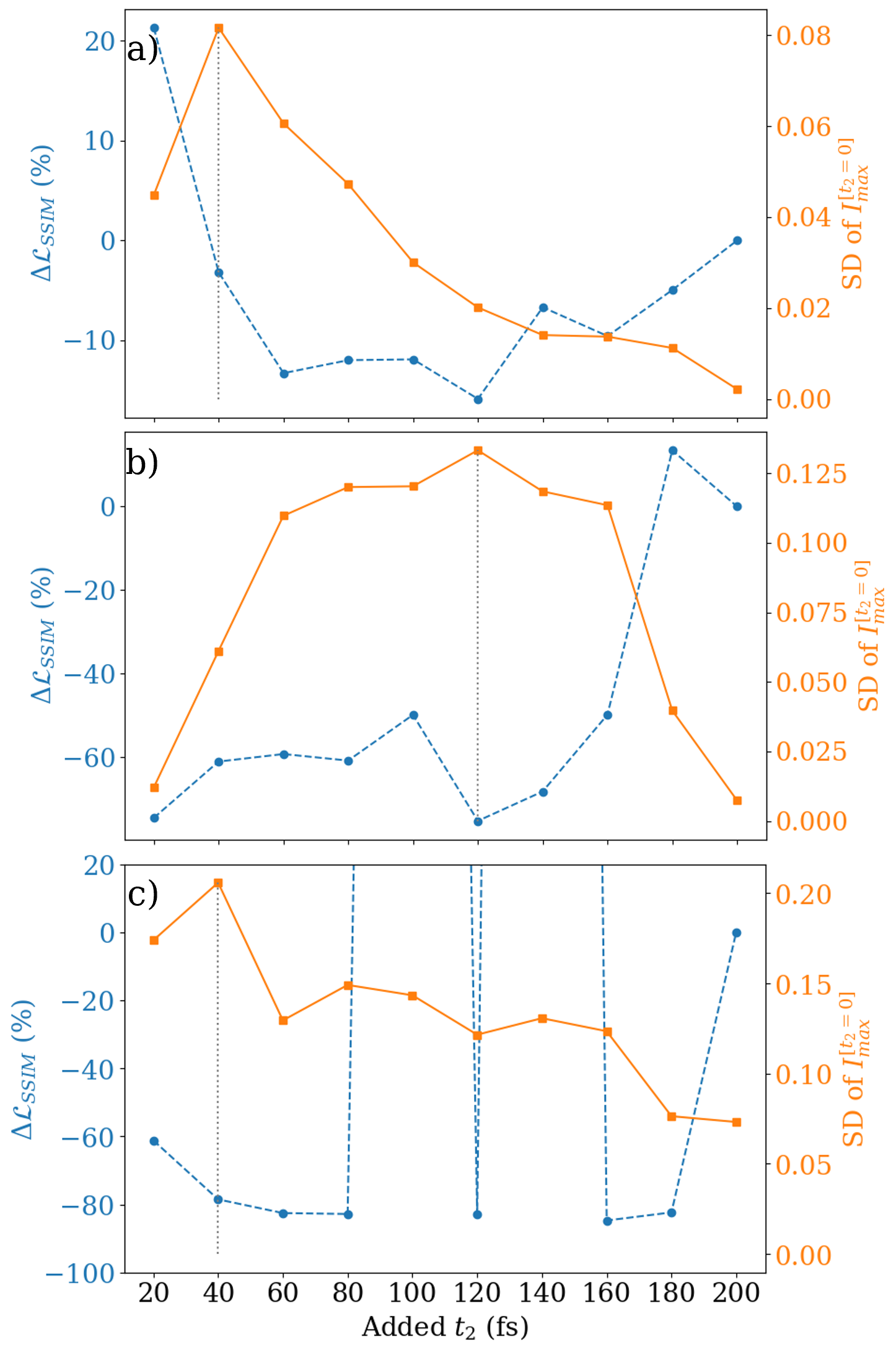}
    \caption{The change in loss of GMM 2DES predictions when incorporating a second $t_2$ time delay into the $t_2$=200~fs fitting procedure. The blue circles (left axis) correspond to the percentage change in mean 2DES loss across the shown $t_2$ range when compared to the $t_2$=200~fs fit with the $t_2$ time delay on the bottom axis incorporated into fitting. The orange squares (right axis) show the SD of the committee $\etmie{0}{t_2}$ predictions. The black dotted lines indicate the $t_2$ time delay where the standard deviation of the $\etmie{0}{t_2}$ is maximal for (a) anionic GFP in water, (b) Nile red in benzene, and (c) PYP in the gas phase.}
    \label{fig:qbc}
\end{figure}

To select the next $t_2$ we computed the standard deviation (SD) of $\etmie{0}{t_2}$ and chose the $t_2$ with the maximum SD as the next training time delay using $M$=10 committee members. This resulted in Fig.~\ref{fig:uncertain}(a, b, c) for anionic GFP in water, Nile red in benzene, and PYP in gas phase, respectively, after training on their corresponding $t_2$=200~fs reference spectrum. Figure~\ref{fig:qbc} shows the change in the SSIM loss relative to the $t_2$=200~fs fit after incorporating the $t_2$ time delay on the x-axis. In general, the $t_2$ time delay using our QbC approach is among those that decrease the loss the most.

\alttext{Line plots of the mean predicted time-dependent maximum intensity with shaded plus or minus 1 standard deviation from a ten-member query-by-committee GMM ensemble for anionic GFP in water, Nile red in benzene, and PYP in the gas phase, each trained on its t2 = 200 femtosecond spectrum. A dashed vertical line marks the time delay of maximum committee uncertainty used for active-learning selection.}

\alttext{Line plots of the percentage change in mean 2DES loss and the standard deviation of the committee's maximum-intensity predictions as a function of the added second time delay, for anionic GFP in water, Nile red in benzene, and PYP in the gas phase. The time delay selected by the query-by-committee approach is generally among those that decrease the loss the most.}

\section{The Effect of the Pulse Spectral Profile on the GMM Fit to Simulated Data}
\label{si:psp_on_nileblue}
\begin{figure}[H]
    \centering
    \includegraphics[width=1\linewidth]{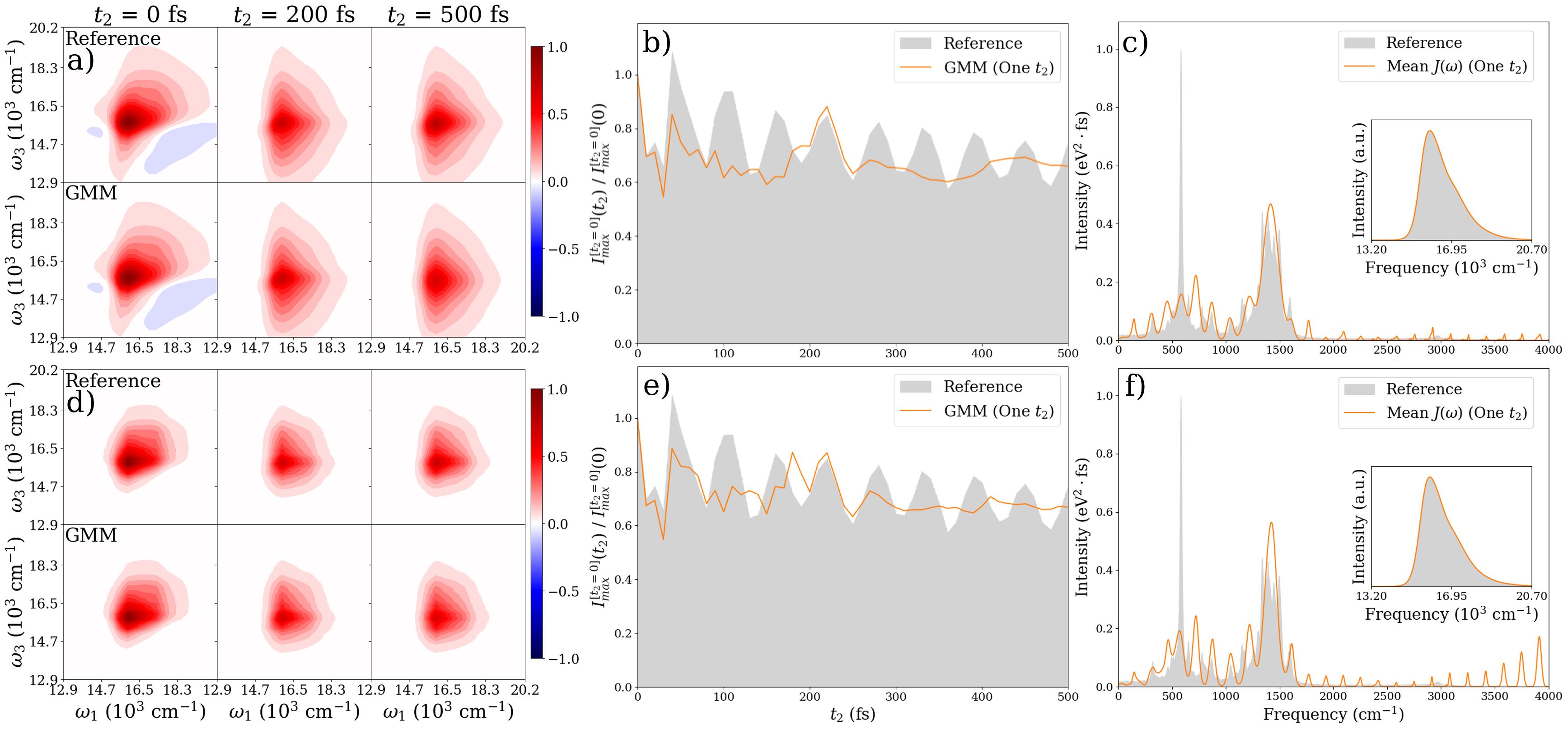}
    \caption{The impact of the pulse spectral profile on GMM fits to simulated Nile blue in ethanol. Top: GMM fits to simulated data of Nile blue in ethanol without the pulse spectral profile. Bottom: GMM fits to simulated data of Nile blue in ethanol with the experimental pulse spectral profile applied to both the reference data and the GMM predicted 2DES during training. (a,d) GMM and reference 2DES spectra. (b,e) GMM and reference $\etmie{0}{t_2}$. (c,f) GMM and reference spectral densities and linear absorption spectra (inset).}
    \label{fig:nb_pulse_effect}
\end{figure}

Here we compare our GMM performance when fitting simulated 2DES for Nile blue in ethanol with and without the experimental pulse spectral profile applied. The top of Fig.~\ref{fig:nb_pulse_effect} shows the GMM fit to the $t_2$=200~fs spectrum without the pulse spectral profile applied. The bottom shows the GMM fit to the same time delay when both the reference 2DES and GMM predicted 2DES have the experimental pulse spectral profile applied. In both cases our method to accurately captures the 2DES, the $\etmie{0}{t_2}$, the features of the reference spectral density, and the linear absorption spectrum.

\alttext{2DES spectra, time-dependent maximum intensity, and spectral density with linear-absorption inset for simulated Nile blue in ethanol fit with and without the experimental pulse spectral profile applied. The GMM accurately recovers the 2DES, intensity decay, spectral density, and linear absorption spectrum in both cases.}

\section{Obtaining the Average Energy Gap from the Linear Absorption Spectrum}
\label{si:wegav_from_linabs}
Here we show that the average energy gap $\weg$ for a two-electronic-level system is equivalent to the weighted average position of the linear absorption spectrum. We used this to initialize our GMM $\weg$ parameter (SI Sec.~\ref{si:gmm_init}). Under the second-order cumulant theoretical framework, the linear absorption spectrum is defined as~\cite{mukamelPrinciplesNonlinearOptical1995,choTwoDimensionalOpticalSpectroscopy2009}
\begin{align*}
\sigma(\omega)=&\text{Re}\int_0^\infty dt |\mu_{eg}|^2e^{-i\weg -g(t)}e^{i\omega t}
\\
=&\frac{|\mu_{eg}|^2}{2}\int_0^\infty dt \left(e^{-i\weg -g(t)}e^{i\omega t}+e^{i\weg -g^*(t)}e^{-i\omega t}\right)
\end{align*}
with the line shape function is given by
\[
g(t)=\int_0^t d\tau_2\int_0^{\tau_2}d\tau_1\langle\delta U(\tau_1)\delta U(0)\rangle.
\]
Taking the weighted average of $\sigma(\omega)$, one obtains
\begin{align}
    \frac{\int_0^\infty d\omega\;\omega\sigma(\omega)}{\int_0^\infty d\omega \sigma(\omega)}=&\frac{\int_0^\infty d\omega\;\omega\frac{|\mu_{eg}|^2}{2}\int_0^\infty dt \left(e^{-i\weg t-g(t)}e^{i\omega t}+e^{i\weg t-g^*(t)}e^{-i\omega t}\right)}{\int_0^\infty d\omega \frac{|\mu_{eg}|^2}{2}\int_0^\infty dt \;e^{-i\weg t-g(t)}e^{i\omega t}+e^{i\weg t-g^*(t)}e^{-i\omega t}}
    \nonumber \\\nonumber\\
    =&\frac{\int_0^\infty d\omega\;\omega\int_0^\infty dt \left(e^{-i\weg t-g(t)}e^{i\omega t}+e^{i\weg t-g^*(t)}e^{-i\omega t}\right)}{\int_0^\infty d\omega \int_0^\infty dt \;e^{-i\weg t-g(t)}e^{i\omega t}+e^{i\weg t-g^*(t)}e^{-i\omega t}}
    \nonumber \\\nonumber\\
    =&\frac{\int_0^\infty dt\int_0^\infty d\omega\;\omega \left(e^{-i\weg t-g(t)}e^{i\omega t}+e^{i\weg t-g^*(t)}e^{-i\omega t}\right)}{\int_0^\infty dt\int_0^\infty d\omega \;e^{-i\weg t-g(t)}e^{i\omega t}+e^{i\weg t-g^*(t)}e^{-i\omega t}}
    \nonumber \\\nonumber\\
    =&\frac{\int_0^\infty dt\int_0^\infty d\omega \;\omega e^{-i\weg t-g(t)}e^{i\omega t}+\omega e^{i\weg t-g^*(t)}e^{-i\omega t}}{\int_0^\infty dt\int_0^\infty d\omega \;e^{-i\weg t-g(t)}e^{i\omega t}+e^{i\weg t-g^*(t)}e^{-i\omega t}}
    \nonumber \\\nonumber\\
    =&\frac{\int_0^\infty dt\int_0^\infty d\omega \;e^{-i\weg t-g(t)}(-i)\frac{d}{dt}\left(e^{i\omega t}\right)+\omega e^{i\weg t-g^*(t)}(i)\frac{d}{dt}\left(e^{-i\omega t}\right)}{\int_0^\infty dt\int_0^\infty d\omega \;e^{-i\weg t-g(t)}e^{i\omega t}+e^{i\weg t-g^*(t)}e^{-i\omega t}}
    \nonumber \\\nonumber\\
    =&\frac{\int_0^\infty dt\;-ie^{-i\weg t-g(t)}\frac{d}{dt}\delta(t)+i e^{i\weg t-g^*(t)}\frac{d}{dt}\delta(t)}{\int_0^\infty dt \;e^{-i\weg t-g(t)}\delta(t)+e^{i\weg t-g^*(t)}\delta(t)}.
\end{align}
The integral in the denominator evaluates to $e^{-g(0)}+e^{-g^*(0)}=2$ because $g(0)=g^*(0)=0$. Now focusing on the numerator, we make use of the fact that
\begin{align}
    \frac{d}{dt}\left[e^{\pm i\weg t-g(t)}\delta(t)\right]=\left[\pm i\weg -g\prime(t)\right]e^{\pm i\weg t-g(t)}\delta(t)+e^{\pm i\weg t-g(t)}\frac{d}{dt}\delta(t).
\end{align}
This can be rearranged and substituted into the expression along with the resolved denominator, giving the weighted average as
\begin{align}
     \frac{\int_0^\infty d\omega\;\omega\sigma(\omega)}{\int_0^\infty d\omega \sigma(\omega)}=&\frac{1}{2}\int_0^\infty dt\;-i\left\{\frac{d}{dt}\left[e^{-i\weg t-g(t)}\delta(t)\right]-\left[-i\weg -g^\prime(t)\right]e^{-i\weg t-g(t)}\delta(t)\right\}
     \nonumber\\&\hspace{1.45cm}+i\left\{\frac{d}{dt}\left[e^{i\weg t-g^*(t)}\delta(t)\right]-\left[i\weg -g^{*\prime}(t)\right]e^{i\weg t-g^*(t)}\delta(t)\right\}
     \nonumber \\\nonumber \\
     =&\frac{1}{2}\left(-i\left\{\left[e^{-i\weg t-g(t)}\delta(t)\right]_0^\infty+i\weg +g^\prime(0)\right\}
     \right.\nonumber\\&\hspace{0.5cm}\left.+i\left\{\left[e^{i\weg t-g^*(t)}\delta(t)\right]_0^\infty-i\weg +g^{*\prime}(0)\right\}\right)
     \nonumber \\\nonumber \\
     =&\frac{1}{2}\left(-i\left\{i\weg +g^\prime(0)\right\}+i\left\{-i\weg +g^{*\prime}(0)\right\}\right)
     \nonumber \\\nonumber \\
     =&\frac{1}{2}\left(\weg -ig^\prime(0)+\weg +ig^{*\prime}(0)\right)
     \nonumber \\\nonumber \\
     =&\weg -i\frac{g^\prime(0)-g^{*\prime}(0)}{2}
     \nonumber \\\nonumber \\
      =&\weg +\mathrm{Im}\left[g^\prime(0)\right].
\end{align}
Since $g^\prime(0)=0$, the weighted average is finally given by $\weg$.

\begin{figure}[H]
    \centering
    \includegraphics[width=1\linewidth]{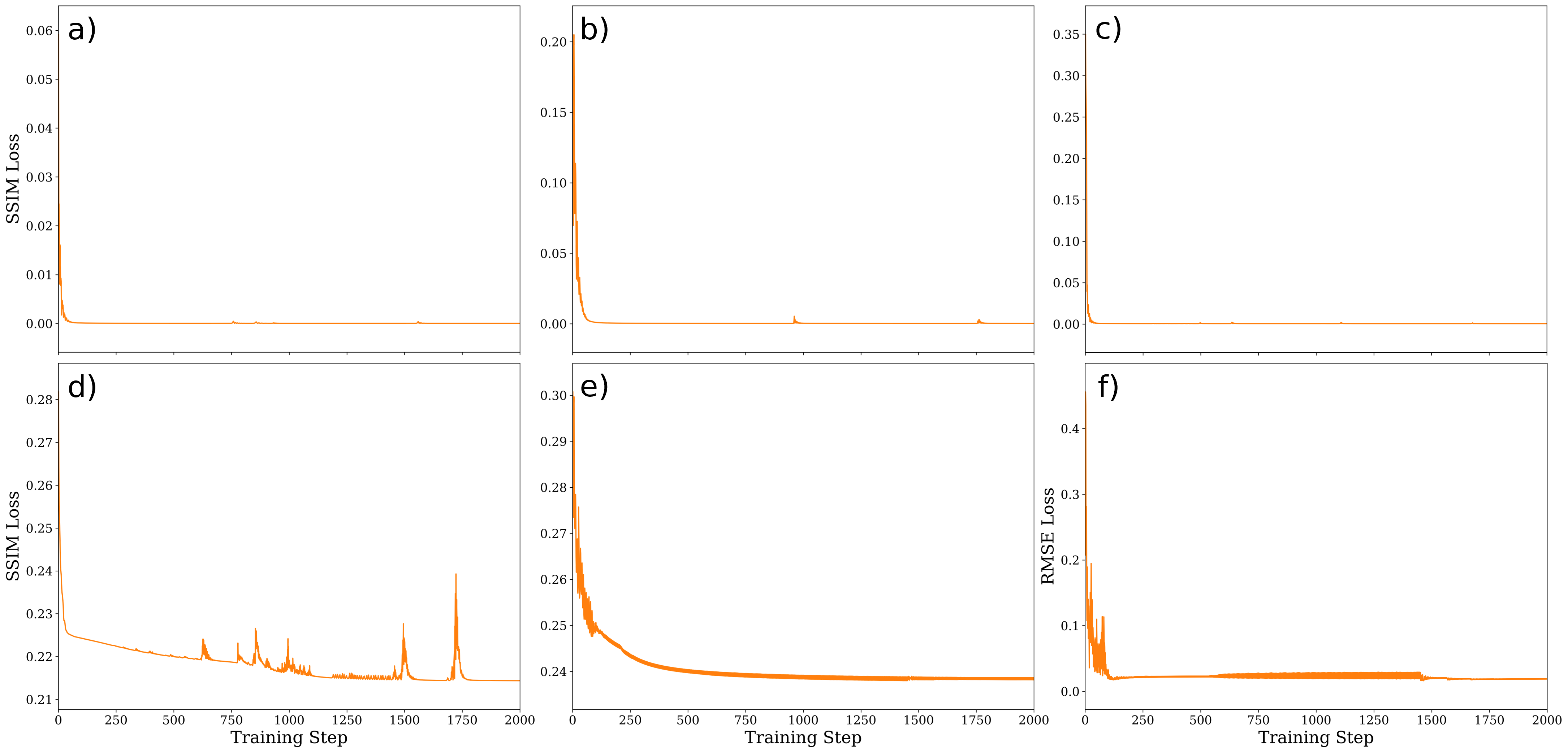}
    \caption{\newchange{Loss vs number of training steps for 2DES predictions (a-e) and linear absorption predictions (f) over the full 2000 epochs of training for GFP in water. The 2DES loss curves shown are for when fitting (a) GFP in water, (b) Nile red in benzene, (c) PYP in vacuum, (d) experimental Nile blue in ethanol with the linear absorption constraint, and (e) experimental Nile blue in ethanol with the linear absorption constraint. (f) The loss curve when fitting the experimental Nile blue in ethanol linear absorption spectrum when fit alongside the 2DES.}}
    \label{fig:loss_curves}
\end{figure}

\newchange{
\section{Loss Curves}
Here we show plots of the loss vs number of training steps for each of the systems discussed in the main text, shown in Fig.~\ref{fig:loss_curves}. Panels (a-e) show the losses of the GMM-predicted 2DES with respect to the reference 2DES, while panel (f) depicts the losses of the GMM-predicted linear absorption spectrum with respect to the reference spectrum.
}
\alttext{Six-panel plot of training loss versus training step for the GMM fits for 2000 training steps. Panels A to E show SSIM loss for the 2DES fits to GFP in water, Nile red in benzene, PYP in vacuum, and experimental Nile blue in ethanol (with and without the linear-absorption constraint), while panel F shows RMSE loss for the linear-absorption fit to Nile blue. All curves drop sharply within the first ~100 steps and then plateau, with the experimental Nile blue fits showing intermittent spikes before settling.}


%

\end{document}